\documentclass[conference]{IEEEtran}
\IEEEoverridecommandlockouts
\usepackage{marvosym}

\usepackage{cite}

\usepackage{tikz}
\usepackage{url}
\usepackage[hidelinks]{hyperref}
\usepackage{amsmath,amssymb,amsfonts}
\usepackage{graphicx}
\usepackage{textcomp}
\usepackage{xcolor}
\usepackage{color}
\usepackage{array}
\usepackage{pifont}
\usepackage{cleveref}
\usepackage{colortbl}
\usepackage{subfigure}
\graphicspath{{./figure/}}
\usepackage{enumitem}
\usepackage[T1]{fontenc}

\usepackage{breakurl}
\usepackage[mathscr]{eucal}
\usepackage{algorithm}
\usepackage{algpseudocode}
\usepackage{xspace}
\usepackage{multirow}
\usepackage{makecell}
\usepackage{booktabs}
\usepackage{diagbox}
\usepackage{threeparttable}
\usepackage{graphicx}
\usepackage{float}
\usepackage{listings}

\usepackage{verbatim}
\usepackage{tcolorbox}
\tcbuselibrary{breakable}
\usepackage{siunitx}
\usepackage{fancyhdr}

\definecolor{JungleGreen}{rgb}{0.16, 0.67, 0.53}
\definecolor{NavyBlue}{rgb}{0, 0.431, 0.722}

\newcommand{\alias}{\texttt{ToolPulse}\xspace}

\begin{document}
\title{\alias: Side-Channel Eavesdropping on Tool-Using LLM Agents}


%\author{
%{\rm Anonymous Author(s)}\\
%Affiliation withheld for review
%}

\author{
\IEEEauthorblockN{
Zihao Dan\IEEEauthorrefmark{1},
Jinyu Zhang\IEEEauthorrefmark{1},
Jinzhao Lian\IEEEauthorrefmark{2},
Yun Chen\IEEEauthorrefmark{3},
Shinan Liu\IEEEauthorrefmark{1}\Letter \thanks{\Letter   Corresponding author}
}

\IEEEauthorblockA{
\IEEEauthorrefmark{1}\textit{The University of Hong Kong}
\quad
\IEEEauthorrefmark{2}\textit{Renmin University of China}\\
\IEEEauthorrefmark{3}\textit{The Hong Kong University of Science and Technology (Guangzhou)}
\quad
}

\IEEEauthorblockA{\footnotesize
u3014019@connect.hku.hk,
u3629294@connect.hku.hk,
2023201891@ruc.edu.cn,
yunchen@hkust-gz.edu.cn,
shinan6@hku.hk
}}


% compsoc conferences compared to traditional conferences), use this
%Telephone: (800) 555--1212, Fax: (888) 555--1212}


\maketitle

\begin{abstract}
Large language models (LLMs) are increasingly deployed as agents that translate user requests into sequences of local tool invocations. While prior cache-based side-channel attacks target the locally deployed models, they typically require the adversary to share model-related components or hardware resources with the victim's inference process. These assumptions do not apply to the predominant agentic settings, in which commercial LLMs execute on vendor-controlled infrastructure. We show that agentic execution exposes a more practical attack surface: although inference remains remote, tools are executed on the user's machine and commonly rely on shared system libraries.
In this paper,  we first systematically evaluate \emph{Flush}+\emph{Reload} and \emph{Prime}+\emph{Probe} for monitoring agentic tool invocations. 
Our evaluation across three hardware platforms shows that \emph{Flush}+\emph{Reload} enables accurate, real-time tool monitoring, whereas \emph{Prime}+\emph{Probe}, despite supporting cross-VM attacks, incurs prohibitive monitoring overhead and suffers from cache-set aliasing.
We present \alias, a cache-based side-channel eavesdropping framework against tool-using agents. We first design the \emph{Tool Recovery Attack} to reconstruct an agent’s tool-execution workflow. We evaluate the \emph{Tool Recovery Attack} on 11 LLMs across Codex and Claude Code. It consistently reconstructs tool workflows, with the lowest F1 score exceeding 93\%. We further demonstrate two attack scenarios enabled by the recovered traces. \emph{Task Profiling Attack} infers user intent across two Terminal-Bench task categories and the \emph{Trigger Localization Attack} identifies sensitive tool calls in encryption/decryption and model training/inference workloads. Even under substantial system noise on a server shared by approximately 20 users, both \emph{Task Profiling Attack} and \emph{Trigger Localization Attack} maintain high accuracy. %\emph{Task Profiling Attack} infers user intent across two Terminal-Bench task categories with accuracies of 87.56\% and 86.07\%, respectively. The \emph{Trigger Localization Attack} identifies sensitive tool calls in encryption/decryption and model training/inference workloads, reducing attack overhead by 75.27\% and 77.09\%, respectively, while maintaining recall above 96\%.
\end{abstract}


% \fi
\IEEEpeerreviewmaketitle

\section{Introduction}
Large language models (LLMs) have rapidly evolved from conversational assistants into practical tools for research and software development. Researchers and developers increasingly rely on LLMs to understand \cite{liu2024repobench} and summarize \cite{sun2025source} codebases, implement code \cite{bairi2024codeplan,yang2024swe}, configure development environments \cite{tang2026devops} and automate complex workflows \cite{shen2024taskbench}.

Recent studies have demonstrated that software isolation alone is insufficient to protect LLM workloads from cache-based side channels. \emph{MoEcho} \cite{ding2025moecho} reveals that the input-dependent expert activation patterns of Mixture-of-Experts (MoE) models leave observable architectural footprints. \cite{gao2025iknow} exploits cache-access patterns of the token-embedding table to reconstruct the inputs and outputs of a locally deployed LLM. %TDXRay \cite{hornetz2026tdxray} attacks LLMs running inside Intel Trust Domain Extensions (TDX) by constructing a hash-map-based mapping from prompts to tokens. 
Collectively, these studies establish that LLM inference exposes secret-dependent hardware behaviors that can be exploited to compromise user privacy. %\textcolor{red}{(I think the red part may not fit smoothly here and might be better placed in the Related Work section) In addition, software-cache-based side-channel attacks have also been explored. For example, PROMPTPEEK \cite{wu2025know} and \cite{song2025early} exploit shared KV caches and Time to First Token (TTFT) to attack users sharing the same LLM serving system.}

\begin{figure*}[tb]
	\centerline{\includegraphics[width=18cm]{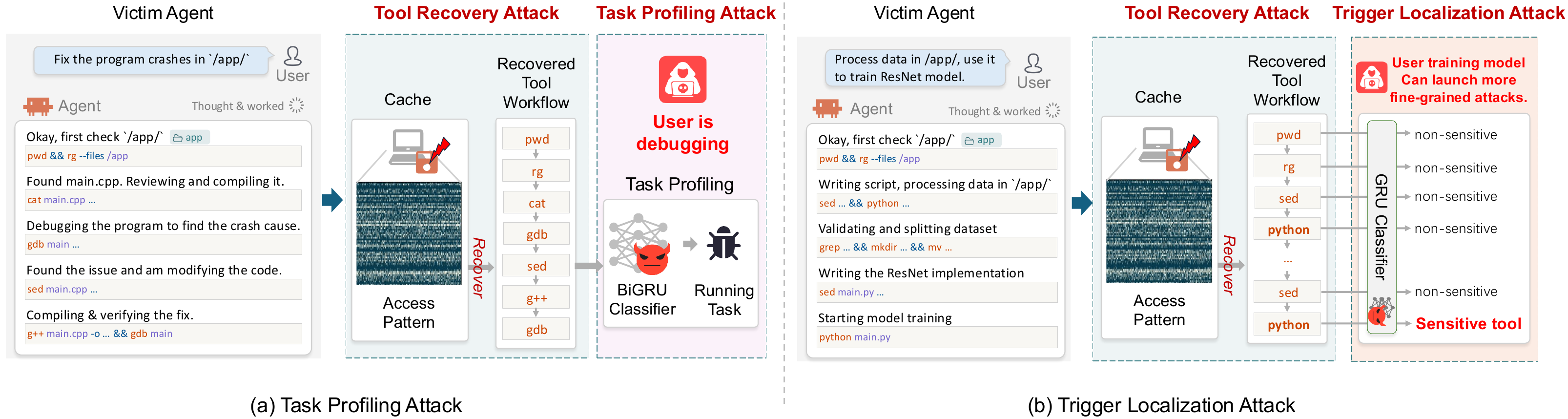}}
	\caption{The Overview of \alias.}
	\label{figure:overview}
\end{figure*}

Despite their significance, existing attacks leave several important gaps. First, they fundamentally assume that the attacker is co-located with the hardware executing the victim model. However, this assumption is unrealistic for the predominant mode of LLM usage. State-of-the-art commercial models, such as GPT \cite{achiam2023gpt}, Gemini \cite{comanici2025gemini}, and Claude \cite{Claude}, are hosted on their vendors' internal infrastructure and accessed by users through web interfaces or APIs. Only a small fraction of users can locally deploy LLMs with comparable capabilities. Consequently, an attacker co-located with the user is unlikely to gain access to the vendor-controlled inference servers, let alone share hardware resources with the victim's LLM instance. Second, prior attacks assume that the adversary can access and share model-related components with the victim, such as experts or embedding tables~\cite{gao2025iknow}. This implies that the adversary must be able to access the victim's model files. In most practical scenarios, however, such access is unrealistic for an unprivileged adversary. %\textcolor{red}{For software-cache-based attacks, the adversary and victim must share the same LLM serving system. This either requires them to use the same API endpoint or requires the adversary to be co-scheduled with the victim on the same backend serving instance, which is difficult to guarantee in practice. Moreover, \cite{luo2025shadow} has shown that the enormous search space required by such attacks makes them impractical.}
These restrictive assumptions substantially limit the practicality of existing side-channel attacks against LLMs. %\shinan{RECONSIDER: the cost of continuous side channel attacks? because they are autoregressive, they may accumulate a lot of previous KV to inspect, and can not start the attack in the middle.}

However, the increasing adoption of agentic LLM deployments~\cite{ClaudeCode,codex,cursor,openhands} may broaden the attack surface. Instead of producing only text, these systems translate user intent into sequences of concrete actions, such as file operations, shell commands, API calls, and database queries. These tool invocations are executed locally on the user’s server and commonly rely on shared libraries provided by the operating system. Consequently, the adversary no longer needs to compromise or gain access to the model vendor’s internal servers. Moreover, the use of shared libraries naturally satisfies the shared-memory requirement of cache-based side-channel attacks, resulting in a practical threat model.
In this paper, we systematically analyze the efficiency of two mainstream cache-based side-channel attacks, \emph{Flush}+\emph{Reload} \cite{yarom2014flushreload} and \emph{Prime}+\emph{Probe} \cite{liu2015last}, for monitoring agentic tool invocations. For \emph{Flush}+\emph{Reload}, a single cache line may contain multiple instructions from the tool executable. Some of these instructions may reside in loops or be revisited due to control-flow branches, causing a single tool invocation to generate multiple cache hits and thus false positives. Conversely, under certain execution paths, the monitored cache line may not be accessed, resulting in false negatives. To address these challenges, we design the Cache Line Localization Algorithm and Deduplication Algorithm. Across three hardware platforms, our approach achieves high F1 scores for monitoring individual tool invocations. We further design an Offset Search Algorithm to infer the arguments passed by the agent to invoked tools. In contrast, although \emph{Prime}+\emph{Probe} enables cross-VM attacks, its high monitoring overhead, cache-set aliasing, and lack of real-time recovery make it impractical for end-to-end attacks.

%we show that the local execution of agent tools creates a practical and powerful side channel. Our key insight is that tool invocations correspond to specific code paths in shared libraries, and these code paths leave distinguishable footprints in the CPU cache. When an attacker shares the same physical machine and library pages with the victim, the attacker can use fine-grained cache monitoring to detect which tools are executed and reconstruct their invocation sequence. This yields a compact representation of the victim's workflow.

Guided by this insight, we present \alias, a cache-based side-channel eavesdropping framework targeting tool-using agents. Figure \ref{figure:overview} provides an overview of \alias. In this framework, we first implement the \emph{Tool Recovery Attack} that can simultaneously monitor multiple tools and output the tool-execution workflow. We evaluate the \emph{Tool Recovery Attack} on 11 models across two agents. The attack achieves consistently high performance across all evaluated models, with the lowest F1 score reaching 93\%. To mitigate noise from background processes and manual tool invocations, we analyzed agent behavior and found that agents tend to invoke tools in  bursts with bash scripts. We therefore record the timestamp of each observed invocation alongside the tool identity, allowing downstream attack models to learn these temporal patterns.

Beyond demonstrating the attack, we further investigate what private information can be leaked through the \emph{Tool Recovery Attack}. To this end, we implement \emph{Tool Recovery Attack} in two attack scenarios: \emph{Task Profiling Attack} and \emph{Trigger Localization Attack}. %Figure \ref{figure:overview} provides an overview of \emph{Task Profiling Attack} and \emph{Trigger Localization Attack}.
\emph{Task Profiling Attack} infers the user’s intent from the tool-execution trace recovered by the \emph{Tool Recovery Attack}. We evaluate its effectiveness on nine task categories from Terminal-Bench 1 and 2 \cite{merrill2026terminal}, achieving classification accuracies of 87.77\% and 91.39\%, respectively. \emph{Trigger Localization Attack} helps the attacker identify when sensitive tools are executed, thereby enabling more fine-grained attacks. We evaluate its effectiveness in two cases, Key Extraction and Model Extraction. While keeping the recall above 96\%, \emph{Trigger Localization Attack} reduces the attack overhead by 80.46\% and 85.79\% for the two case studies, respectively. To assess the impact of user-generated tool-level noise on \emph{Task Profiling Attack} and \emph{Trigger Localization Attack}, we replayed tool invocations recorded on a server shared by approximately 20 users in a controlled environment. Despite the substantial user-generated noise, the accuracy of \emph{Task Profiling Attack} drops by only about 4\%, and the recall of \emph{Trigger Localization Attack} drops by about 1-3\%, indicating that \alias is robust to such noise.
%\shinan{worth mentioning that we can target Codex, Claude Code, OpenHands these agents}

We make the following contributions:

\begin{itemize}[leftmargin=*]
    \item \textbf{We identify tool execution as a practical side-channel attack surface for LLM agents.} An adversary co-located with the agent can eavesdrop on local tools via shared microarchitectural resources, even if the LLM runs remotely.

    \item \textbf{We systematically characterize cache side channels for recovering tool invocations.} We evaluate \textsc{Flush+Reload} and \textsc{Prime+Probe} techniques. Cloud CPU tests show \textsc{Flush+Reload} recovers diverse tool invocations with up to 100\% F1 scores.

    \item \textbf{We design an end-to-end attack to reconstruct LLM-agent workflows.} By monitoring multiple tools, an unprivileged adversary can rebuild tool sequences without ever observing the model's prompts or computation.

    \item \textbf{We demonstrate that recovered workflows expose critical high-level agent activity.} We can classify agent tasks in the two datasets with accuracies of 87.77\% and 91.39\%, respectively. We also introduce the \emph{Trigger Localization Attack}  that flags security-sensitive operations in real-time to trigger targeted, fine-grained monitoring.
\end{itemize}

%\noindent \textbf{Responsible Disclosure:} We responsibly disclosed \alias to the OpenAI and Anthropic Security Team.
%\noindent\textbf{Artifact.} The source code of this work is available at https://anonymous.4open.science/r/ToolPulse-861B/ and we will release all datasets used for training and evaluation upon acceptance of the paper.
% The emergence of agents has liberated large language models (LLMs) from the traditional question-and-answer paradigm, enabling them to perform more sophisticated interactive tasks such as code generation, web browsing, and computer use. These capabilities are largely enabled by external tools that LLMs can invoke. This ecosystem originated from function calling introduced by OpenAI, and has since evolved to include standardized interfaces for external tools (e.g., the MCP protocol) as well as abstractions such as skills that facilitate better reasoning over tool usage. Today, a rich agent ecosystem has significantly enhanced the ability of LLMs to interact with their environments efficiently while minimizing token overhead.

% However, this frequent invocation of external tools raises an important question: does it introduce new risks for user privacy leakage?
\noindent \textbf{Responsible Disclosure:} We responsibly disclosed \alias to the Openai and Anthropic Security Team on September 21.
\section{Background and Related Work}
In this section, we provide background on cache side-channel attacks and LLM agents. We also discuss existing side-channel attacks against LLMs.
\subsection{Cache Side-channel Attacks}
Modern processors employ caches to bridge the gap between their high computational speed and the relatively slow access latency of main memory \cite{smith1982cache}. A cache is a small, low-latency storage structure located closer to the processor than DRAM. Although its capacity is significantly smaller than that of main memory, it provides much faster data access \cite{kowarschik2003overview}.

Leveraging the principle of temporal locality \cite{snir2005theory}, processors load recently accessed data into the cache, as such data is likely to be accessed again in the near future. Consequently, subsequent accesses can be served with substantially lower latency. However, this optimization also introduces a side channel: by measuring memory access latency, an attacker can determine whether a particular memory location resides in the cache and thereby infer the victim’s access patterns. This class of attacks \cite{liu2015last,zhao2024last,yan2019attack,gruss2016flush+,yarom2014flushreload} is known as cache side-channel attacks. Cache side-channel attacks can be broadly categorized into two classes: shared-based attacks and contention-based attacks.

\textbf{Shared-Memory Attacks.} Shared-memory attacks require the adversary and the victim to access the same physical memory locations. Typical settings include shared libraries~\cite{gruss2016flush+} and memory deduplication~\cite{arcangeli2009increasing}. This class of attacks was popularized by \emph{Flush}+\emph{Reload}, which leverages the \texttt{CLFLUSH} instruction to evict a shared memory line from the cache. If the victim subsequently accesses the same line, the line is brought back into the cache. By measuring whether selected shared-memory lines are cached after the victim’s execution, the adversary can infer the victim’s access pattern. Since \emph{Flush}+\emph{Reload}~\cite{yarom2014flushreload}, several variants have been proposed. For example, \emph{Flush}+\emph{Flush}~\cite{gruss2016flush+} improves temporal resolution by avoiding reloads, while \emph{Evict}+\emph{Reload} removes the \texttt{CLFLUSH} dependency to evict target cache lines.

\textbf{Contention-Based Attacks.} Contention-based attacks exploit interference in shared microarchitectural resources and are commonly represented by \emph{Prime}+\emph{Probe}~\cite{liu2015last}. In a \emph{Prime}+\emph{Probe} attack, the adversary first constructs an eviction set that maps to the same cache set as the victim’s target address and uses it to prime the cache. If the victim accesses data mapping to that cache set, it evicts one or more of the adversary’s cache lines. The adversary then probes the eviction set and observes increased access latency, from which the victim’s cache-set usage can be inferred.

\subsection{LLM Agent}
LLM agents extend conventional language models with the ability to interact with external tools and execution environments \cite{kim2026sok}. Rather than merely generating textual responses, an agent exposes a set of tools to the LLM through structured interfaces, such as shell commands \cite{yang2024swe}, file-system operations \cite{xie2024osworld}, web browsers \cite{deng2023mind2web}, and application programming interfaces (APIs) \cite{qin2024toolllm}. The LLM selects a tool and generates its arguments based on the user request, while the agent runtime validates and executes the resulting tool call and returns its output to the LLM. This iterative reasoning and execution loop allows the model to adapt subsequent actions according to intermediate results \cite{huang2024understanding}. By bridging language-model reasoning with external execution, agents transform LLMs from passive chatbots into autonomous systems capable of performing real-world tasks, including writing/testing code, inspecting repositories, installing software dependencies, configuring environments, and operating external services.

This work focuses on computer-related agentic tasks, including coding, software testing, and environment configuration. Accordingly, we consider tools whose execution relies on system executables and shared libraries whose code pages are commonly shared across users and processes. Web agents and other agent paradigms that rely on different tool implementations are outside the scope of this work.

\subsection{Related Work}

To the best of our knowledge, existing side-channel attacks against LLMs can be broadly categorized into three types: hardware-cache-based side-channel attacks, software-cache-based attacks that exploit the KV cache, and network-traffic-based side-channel attacks.

\textbf{Hardware Cache Attacks.} This category of attacks is the most closely related to our work. Most of them attempt to infer user-sensitive information by monitoring memory accesses associated with the internal execution state of an LLM. For example, \emph{MoEcho} \cite{ding2025moecho} monitors memory accesses corresponding to different experts in an MoE model, while \cite{gao2025iknow} monitors memory accesses associated with different tokens in the model tokenizer. These attacks typically require not only co-location with the server hosting the victim LLM, but also access to the same model files or model-related memory regions as the victim. Moreover, their attack surfaces vary across hardware platforms, since LLM inference may be executed on either CPUs or GPUs. \emph{TDXRay} \cite{hornetz2026tdxray} attacks LLMs running inside Intel Trust Domain Extensions (TDX) by constructing a hash-map-based mapping from prompts to tokens. Although it does not require the attacker and victim to share the same library, the attacker must be able to query the victim LLM to identify the physical memory addresses associated with specific tokens.

\textbf{Software Cache Attacks.} Another category of attacks exploits KV caches that may be shared across users. \emph{PROMPTPEEK} \cite{wu2025know} leverages SGLang’s scheduling policy, which prioritizes inputs with greater overlap with cached KV states, to infer private information from other users. \cite{song2025early} similarly exploits KV-cache and semantic-cache reuse to infer system prompts and user input. However, the adversary and victim must share the same LLM serving system in software-cache-based side-channel attacks. This either requires them to use the same API endpoint or requires the adversary to be co-scheduled with the victim on the same backend serving instance, which is difficult to guarantee in practice. Moreover, \cite{luo2025shadow} has shown that the enormous search space required by such attacks makes them impractical.

\textbf{Network Traffic Attacks.} Another line of side-channel attacks seeks to infer information about user–LLM interactions by analyzing the encrypted network traffic exchanged between LLM services and users. \cite{WeissWhat2024} exploited token-length information leaked through encrypted traffic to leak user task topic and reconstruct LLM responses. \cite{lilength} extended this attack to merged-token streaming, showing that aggregated token lengths remain vulnerable to semantic inference. However, these attacks generally achieve limited effectiveness as the traffic patterns are similar. In the topic-inference setting most comparable to ours, \cite{WeissWhat2024} achieve only a 53\% success rate, while \cite{lilength} report an even lower rate of 42.5\%.

\section{Threat Model}
In this work, consistent with prior cache-based side-channel studies \cite{zhao2024last,liu2015last,yarom2014flushreload,gruss2016flush+}, we assume that the attacker shares the same physical CPU resources with the victim. The attacker and the victim may correspond to two legitimate users sharing the same server, or the attacker may take the form of a malicious process. In addition, prior works \cite{zhang2014cross,zhao2024everywhere} have extensively demonstrated how an adversary can achieve co-location with a victim in public-cloud environments.

We assume that the victim deploys agent tools (e.g. Linux Commands) on a working server, while the underlying model accessed by the victim is hosted remotely by a model provider. As a result, the attacker cannot directly obtain model execution information via traditional side channels. This setting reflects a common real-world deployment scenario. For example, when using Claude Code/Codex \cite{ClaudeCode,codex} on an Ubuntu machine, Linux commands are invoked locally, whereas the actual model execution takes place on the model provider’s infrastructure and communicates with the client via APIs.

For \textit{Flush+Reload}-based side-channel attacks, consistent with prior work, we assume that the attacker and the victim share the same agent tool library. Across all attack scenarios, the attacker operates with unprivileged user permissions and does not possess elevated privileges.
\section{Side Channels in Agentic Systems}
In this section, we systematically analyze  \emph{Flush}+\emph{Reload}~\cite{yarom2014flushreload} and \emph{Prime}+\emph{Probe}~\cite{liu2015last} threats against LLM-agent tools, and evaluate their practicality for enabling an end-to-end attack.
\subsection{\emph{Flush}+\emph{Reload}-based Attacks}
\label{subsection:flush+reload}
We build our \emph{Tool Recovery Attack} on top of the traditional \emph{Flush}+\emph{Reload} techniques \cite{yarom2014flushreload,gruss2015cache,gruss2016flush+}. The attacker targets the code segments of shared libraries invoked during victim LLM tool execution, and continuously monitors tool usage by repeatedly performing the following loop:

\begin{itemize}[leftmargin=*]
    \item \textbf{Flush:} Evict a target address within the code segment from the cache to ensure it is not resident.
    \item \textbf{Wait:} Sleep for a predefined time window to allow the victim to execute.
    \item \textbf{Reload:} Load the same address and measure the latency to determine whether it has been brought back into the cache.
\end{itemize}

\begin{figure}[tb]
	\centerline{\includegraphics[width=8.5cm]{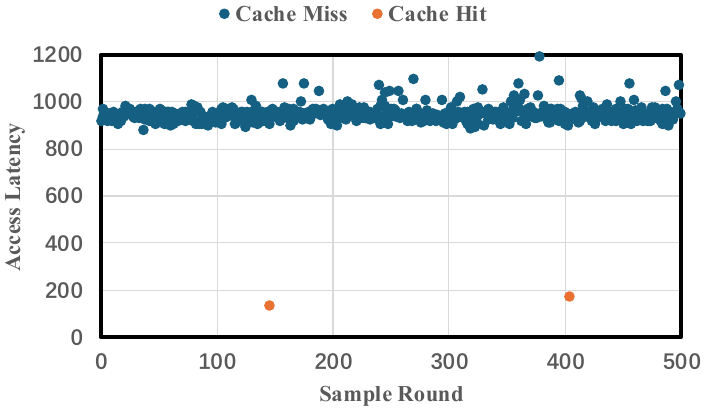}}
	\caption{Results of 500 \emph{Flush}+\emph{Reload} attack rounds for monitoring LLM function calls: the cache hit indicates that the function is invoked by the victim LLM.}
	\label{figure:flushexa}
\end{figure}

Figure \ref{figure:flushexa} illustrates 500 rounds of Flush+Reload measurements targeting a specific library. As highlighted in orange, two samples exhibit significantly lower access latency compared to the others, indicating that the victim accessed the target address during the waiting window and brought the corresponding data into the cache.

\begin{figure}[tb]
	\centering
	\subfigure[Single Invocation with One Cache Hit.]{\includegraphics[width=8.5cm]{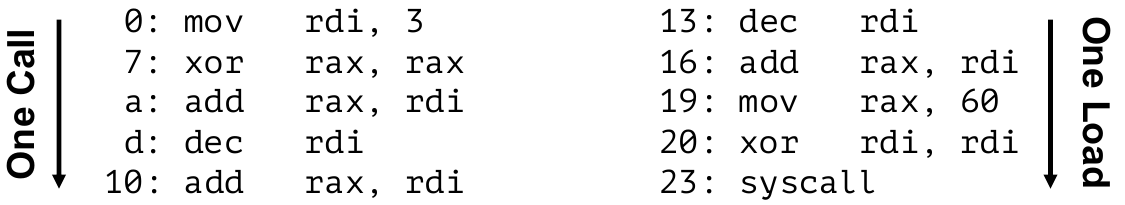}\label{figure:sighit}}
	\subfigure[Single Invocation with Multiple Cache Hits.]{\includegraphics[width=8.5cm]{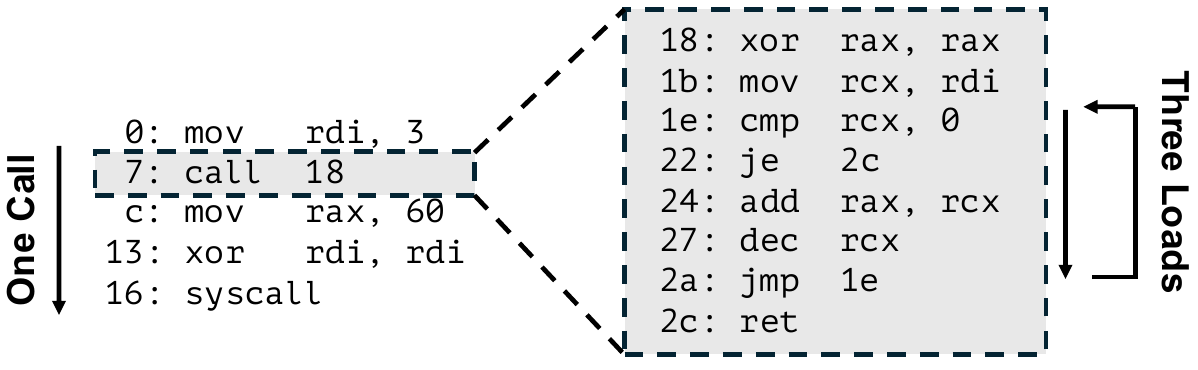}\label{figure:multihit}}
	\caption{Examples illustrating why a single tool invocation could lead to multiple cache hits.}
    \label{figure:assem}
\end{figure}

However, this does not necessarily imply that the victim LLM performed two distinct tool invocations. This is because, during function execution, control flow may repeatedly jump to the same code address multiple times. Figure \ref{figure:assem} further provides simple assembly-level examples to illustrate this behavior. If the code is executed strictly sequentially, as shown in Figure \ref{figure:sighit}, a single invocation will produce only one cache hit. However, although the example in Figure \ref{figure:multihit} issues only a single tool invocation, the code segment from \texttt{0x1e} to \texttt{0x2a} is accessed three times during execution. This can result in multiple cache hits on the same cache line. Moreover, since a cache line spans 64 bytes, it may contain multiple instructions. As a result, function control-flow transitions can cause the same cache line to be loaded multiple times. For example, invoking a function at address \texttt{0x7} and later resuming execution at address \texttt{0xc}. These repeated accesses may introduce \emph{false positives} in the attack. Therefore, the attacker should ideally select the cache lines that are unlikely to be repeatedly accessed.% However, since shared libraries are often distributed as precompiled binaries, the attacker cannot guarantee that a given code offset will not be accessed multiple times.

%\begin{figure}[tb]
%	\centerline{\includegraphics[width=8.5cm]{figure/assemble.pdf}}
%	\caption{Example illustrating why a single tool invocation leads to multiple cache hits.}
%	\label{figure:assem}
%\end{figure}

\textbf{Cache Line Localization.} The selection of cache lines can introduce not only false positives, but also false negatives. Since the attacker does not know the arguments used when the agent invokes a tool, the attacker cannot determine which execution branches will be taken. If all instructions within the selected cache line are skipped under certain argument conditions, the tool invocation will not be detected even though the agent executes the tool, thereby leading to \emph{false negatives}.

Therefore, an ideal cache line should satisfy three properties: 
(1) \texttt{Property 1}: Some instructions within the cache line are accessed along all execution paths of the target tool.
(2) \texttt{Property 2}: The instructions within the cache line are not located inside loops. (3) \texttt{Property 3}: The cache line is accessed only once during each execution path of the tool.

Since the tools shared with the victim agent are typically distributed as compiled executables rather than source code, we employ binary analysis techniques to identify optimal cache lines for \emph{Flush}+\emph{Reload} attacks. Specifically, we use angr \cite{shoshitaishvili2016state} to recover the tool’s control-flow graph (CFG), and design the Cache Line Localization Algorithm (CLLA) to identify the suitable cache line for the attack. The details of CLLA are provided in Algorithm \ref{alg:CLA} in Appendix \ref{appendix:argdetail}.

The core idea of the CLLA is to leverage the CFG recovered by angr to enumerate all feasible execution paths and eliminate cache lines that can be bypassed along any path. This step satisfies \texttt{Property 1} and is relatively easy to achieve in practice. In our experiments, nearly all evaluated tools contain at least one such cache line, ensuring that \emph{Flush}+\emph{Reload} attacks do not suffer from false negatives caused by cache-line selection.

Among the remaining candidates, we first prioritize cache lines containing the fewest instructions located inside loops, and then select the cache line with the lowest average access count across all execution paths. Table \ref{table:locatecache} reports the results for 18 representative tools across different operating system versions. The results show that, for most tools, it is possible to identify cache lines whose instructions are not located within loops. However, for the majority of tools, it is not possible to find a cache line that is accessed exactly once on every execution path. Consequently, even the best cache line identified by our algorithm may still generate false positives due to repeated accesses during a single tool invocation.

\begin{table}[t]
	\caption{Result of Cache Line Localization Algorithm.} 
    \small
	\begin{center}
		\setlength\tabcolsep{3pt}
		\begin{threeparttable}
			\begin{tabular}{c|c|c|c|c|c|c|c|c}
				\toprule[2pt]
                \multirow{2}{*}{Tools} & \multicolumn{2}{c|}{ubuntu22.04} & \multicolumn{2}{c|}{ubuntu26.04} & \multicolumn{2}{c|}{centos7.8} & \multicolumn{2}{c}{debian11.6}\\
                \cline{2-9}
                 & Avg\tnote{1} & Loop\tnote{2} & Avg & Loop & Avg & Loop & Avg & Loop\\
                \hline
                cp & 6 & 0 & 11 & 0 & 11 & 0 & \cellcolor{NavyBlue!30}1 & \cellcolor{NavyBlue!30}0\tnote{3}\\
                \hline
                rm & 8 & 0 & 11 & 0 & \cellcolor{NavyBlue!30}1 & \cellcolor{NavyBlue!30}0 & 11 & 0\\
                \hline
                mkdir & 6 & 0 & 11 & 0 & \cellcolor{NavyBlue!30}1 & \cellcolor{NavyBlue!30}0 & 11 & 0\\
                \hline
                mv & 12 & 0 & \cellcolor{NavyBlue!30}1 & \cellcolor{NavyBlue!30}0 & 3.84 & 3 & 3 & 0\\
                \hline
                bash & 6 & 0 & 3 & 0 & 3.98 & 3 & 11 & 0\\
                \hline
                pwd & 6 & 0 & 11 & 0 & 3.57 & 3 & \cellcolor{NavyBlue!30}1 & \cellcolor{NavyBlue!30}0\\
                \hline
                sed & 8 & 0 & 11 & 0 & 3.48 & 3 & 11 & 0\\
                \hline
                sort & 7.03 & 0 & 11 & 0 & 2.88 & 2 & \cellcolor{NavyBlue!30}1 & \cellcolor{NavyBlue!30}0\\
                \hline
                grep & 6 & 0 & 7 & 0 & 3.92 & 3 & 3 & 0\\
                \hline
                git & 6 & 0 & 12 & 0 & 3.99 & 3 & \cellcolor{NavyBlue!30}1 & \cellcolor{NavyBlue!30}0\\
                \hline
                echo & 5.04 & 0 & 9 & 0 & 3.56 & 3 & 11 & 0\\
                \hline
                file & 12 & 0 & 6 & 0 & 3 & 0 & 11 & 0\\
                \hline
                fasttext & 12 & 0 & 12 & 0 & \cellcolor{NavyBlue!30}1 & \cellcolor{NavyBlue!30}0 & \cellcolor{NavyBlue!30}1 & \cellcolor{NavyBlue!30}0\\
                \hline
                gzip & 7 & 0 & 6 & 0 & 3.65 & 3 & 11 & 0\\
                \hline
                awk & 12 & 0 & 12 & 0 & \cellcolor{NavyBlue!30}1 & \cellcolor{NavyBlue!30}0 & \cellcolor{NavyBlue!30}1 & \cellcolor{NavyBlue!30}0\\
                \hline
                dpkg & 6 & 0 & 12 & 0 & 4 & 4 & 11 & 0\\
                \hline
                make & 7 & 0 & 6 & 0 & \cellcolor{NavyBlue!30}1 & \cellcolor{NavyBlue!30}0 & 11 & 0\\
                \hline
                gcc & 12 & 0 & 8 & 0 & 1 & 1 & 1 & 3\\
				\bottomrule[2pt]
			\end{tabular}
            \begin{tablenotes}
				\item[1] Avg denotes the average number of times the cache line is accessed along each execution path.
				\item[2] Loop denotes the number of instructions within the cache line that are located inside loops.
				\item[3] Cells marked in blue indicate that the corresponding tool has an ideal cache line that doesn't introduce false positives.
			\end{tablenotes}
		\end{threeparttable}
		\label{table:locatecache}
	\end{center}
    \vspace{-0.5cm}
\end{table}

\textbf{Deduplication Attack.} To address the false positives caused by repeated accesses, we design a Deduplication Attack Algorithm (DAA) to filter out redundant accesses and improve the accuracy of trace extraction. Due to the principle of spatial locality, accesses to the same code address tend to be clustered within a short time window. Consequently, a large number of cache misses typically occur between two function calls. Based on this observation, the core idea of DAA is as follows: upon detecting a cache hit, the algorithm does not immediately treat subsequent hits as separate events. Instead, it considers a function call to have ended only after observing a sufficient number of consecutive cache misses. All cache hits within this interval are thus grouped together and treated as multiple accesses to the same code segment within a single function invocation. The details of DAA are provided in Algorithm \ref{alg:da} in Appendix \ref{appendix:argdetail}.

\begin{table}[t]
	\caption{Ablation Study of CLLA and DAA.} 
    \small
	\begin{center}
		\setlength\tabcolsep{3pt}
		\begin{threeparttable}
			\begin{tabular}{c|c|c|c|c}
				\toprule[2pt]
                Cache Line & Attack Method & Recall & Precision & F1\\
                \hline
                \multirow{2}{*}{Without CLLA\tnote{1}} & Without DAA & 22.2\% & 28.1\% & 24.8\% \\
                \cline{2-5}
                 & With DAA &  17.6\% & 98.9\% & 29.9\%\\
                \hline
                \multirow{2}{*}{With CLLA\tnote{2}} & Without DAA & 100\% & 27.3\% & 42.9\% \\
                \cline{2-5}
                 & With DAA &  99.6\% & 100\% & 99.8\%\\
				\bottomrule[2pt]
			\end{tabular}
        \begin{tablenotes}
				\item[1] Without CLLA denotes the setting in which the monitored cache line is bypassed by some execution paths.
				\item[2] With CLLA denotes the setting in which the monitored cache line is selected using CLLA.
		\end{tablenotes}
		\end{threeparttable}
		\label{table:ablation}
	\end{center}
    \vspace{-0.5cm}
\end{table}

\begin{table*}[t]
	\caption{Evaluation of \emph{Flush}+\emph{Reload} in Agentic Tools.} 
    \footnotesize
	\begin{center}
		\setlength\tabcolsep{3.8pt}
		\begin{threeparttable}
			\begin{tabular}{c|c|c|c|c|c|c|c|c|c|c|c|c|c|c|c|c}
				\toprule[2pt]
                \multicolumn{2}{c|}{Type} & \multicolumn{5}{c|}{Linux Command} & \multicolumn{5}{c|}{Git} & \multicolumn{5}{c}{Other Applications}\\
                \hline
                \multicolumn{2}{c|}{Tools} & mv & rm & cp & cd & mkdir & clone & commit & add & diff & status & python & R & gzip & openssl & hexdump\\
                \hline
                \multirow{3}{*}{\makecell{Xeon\\8369}} & Rec & 99.8\% & 100\% & 99.8\% & 92.4\% & 99.6\% & 100\% & 98.8\% & 100\% & 99.6\% & 99.8\% & 100\% & 99.6\% & 100\% & 100\% & 100\%\\
                \cline{2-17}
                & Prec & 97\% & 100\% & 99.8\% & 95.5\% & 100\% & 100\% & 100\% & 100\% & 100\% & 100\% & 100\% & 100\% & 99.8\% & 100\% & 100\%\\
                \cline{2-17}
                & F1 & 98.4\% & 100\% & 99.8\% & 93.9\% & 99.8\% & 100\% & 99.4\% & 100\% & 99.8\% & 99.9\% & 100\% & 99.8\% & 99.9\% & 100\% & 100\%\\
                \hline
                \multirow{3}{*}{\makecell{Xeon\\8269}} & Rec & 98\% & 100\% & 99.8\% & 97.4\% & 99.8\% & 98.8\% & 94.4\% & 96.8\% & 98.2\% & 98.2\% & 100\% & 100\% & 99.4\% & 100\% & 99.8\%\\
                \cline{2-17}
                & Prec & 100\% & 99.8\% & 100\% & 97.2\% & 100\% & 100\% & 99.2\% & 98.2\% & 100\% & 100\% & 100\% & 100\% & 100\% & 100\% & 100\%\\
                \cline{2-17}
                & F1  & 99\% & 99.9\% & 99.9\% & 97.3\% & 99.9\% & 99.4\% & 96.7\% & 97.5\% & 99.1\% & 99.1\% & 100\% & 100\% & 99.7\% & 100\% & 99.9\%\\
                \hline
                \multirow{3}{*}{\makecell{Xeon\\8163}} & Rec & 99.8\% & 99.8\% & 99.8\% & 95\% & 99.6\% & 99.2\% & 95\% & 95.6\% & 97.6\% & 99.8\% & 99.8\% & 98.6\% & 99.4\% & 99.8\% & 99.8\%\\
                \cline{2-17}
                & Prec & 100\% & 100\% & 100\% & 94.6\% & 100\% & 100\% & 100\% & 100\% & 100\% & 100\% & 100\% & 100\% & 100\% & 100\% & 100\%\\
                \cline{2-17}
                & F1  & 99.9\% & 99.9\% & 99.9\% & 94.8\% & 99.8\% & 99.6\% & 97.4\% & 97.6\% & 98.8\% & 99.9\% & 99.9\% & 99.3\% & 99.7\% & 99.9\% & 99.9\%\\
				\bottomrule[2pt]
			\end{tabular}
            %\begin{tablenotes}
			%	\item[1] All values in the table are percentages.
			%\end{tablenotes}
		\end{threeparttable}
		\label{table:asrflushreload}
	\end{center}
    \vspace{-0.5cm}
\end{table*}

In Algorithm \ref{alg:da}, $addr$ denotes the selected cache line address, which is known to the adversary. The parameter $\tau$ represents the latency threshold used to distinguish between cache hits and cache misses in the \emph{Flush}+\emph{Reload} attack; the adversary can calibrate this threshold by measuring the latency of cache hits and misses. The parameter $U$ specifies the number of consecutive cache misses required to determine the end of a function invocation. This hyperparameter depends on several factors, including the CPU microarchitecture of the physical machine, the characteristics of the victim library, and the interval configuration between the \emph{Flush} and \emph{Reload} phases. Therefore, $U$ must be carefully tuned by the adversary for each specific deployment scenario.

\textbf{Ablation Study.} In this section, we investigate the impact of the CLLA and DAA on monitoring agent tools. We use Python as a representative tool to evaluate the effectiveness of both algorithms. We first apply CLLA to identify the optimal cache line for monitoring Python. Similar to most evaluated tools, the average number of accesses to this cache line along an execution path is not exactly one. However, the selected cache line is accessed by all feasible execution paths. For comparison, we additionally select another cache line that is bypassed by some execution paths, which we refer to as the missed cache line. We then evaluate the monitoring performance using both the optimal and missed cache lines, with and without DAA. Table \ref{table:ablation} summarizes the results, allowing us to separately quantify the benefits of CLLA and DAA.

The results show that CLLA primarily reduces false negatives by selecting a cache line that is accessed along all feasible execution paths. However, repeated accesses to the selected cache line can still produce false positives. DAA addresses this issue by deduplicating multiple cache hits originating from the same tool invocation, thereby substantially reducing false positives.

\textbf{Search Target Address Offset.} To achieve more fine-grained attacks, the adversary may aim to infer specific command arguments used in LLM tool invocations. For example, distinguishing whether the LLM executes \texttt{git clone} or \texttt{git commit}. This requires identifying code segments that are uniquely associated with particular command arguments. The selection of such victim addresses must satisfy two conditions: (1) the target address is accessed when the specific command argument is executed, and (2) the address is not accessed during the execution of other command arguments. As illustrated in Figure \ref{figure:assem}, if the branch corresponding to \texttt{call 0x18} is exclusively triggered by \texttt{git clone}, then the code segment from \texttt{0x18} to \texttt{0x2c} is a suitable target for inferring whether the LLM invokes \texttt{git clone}.

However, the adversary cannot directly obtain the offset of such code segment relative to the base address of the shared library. To address this challenge, we design the Offset Search Algorithm (OSA) to automatically search the required code offsets for the attack. In this algorithm, we scan the entire memory range of the target library at a granularity of the cache line size (64 bytes). For each candidate address, we execute all instruction arguments as workloads and perform $R$ rounds of \emph{Flush}+\emph{Reload} measurements. We record the number of cache hits corresponding to the target instruction argument, as well as the aggregated cache hits from all other arguments, and compute a score based on their difference. If the computed score exceeds the current $best\_score$, we update both $best\_score$ and $best\_offset$. If the score reaches $R$, we return the corresponding address, as this represents the maximum achievable score. The implementation details of OSA are provided in Algorithm \ref{alg:offset_search} in Appendix \ref{appendix:argdetail}.

In subsequent experiments, we observe that \textbf{\textit{as long as the same version of the executable is used (e.g., Linux commands under the same OS distribution), the identified code offsets are reusable across different physical machines}}. This is because, for a given binary, the offset of each branch target remains fixed. This observation implies that the adversary does not need to repeat the discovery process on different machines, significantly reducing the attack overhead.

\textbf{Real-world evaluation.} We evaluate our Flush+Reload-based attack on three real-world public cloud instances, equipped with Intel Xeon Platinum 8369B, Intel Xeon Platinum 8269CY, and Intel Xeon Platinum 8163, respectively. 

We evaluate our attack on 15 tool types across three categories that can be invoked by LLMs. For each tool, we conduct 1000 rounds of experiments. In each round, we randomly invoke the tool five times with intervals ranging from 100 ms to 1 s. During each invocation, we randomize the input arguments (e.g., file paths and sizes for \texttt{cp}) to comprehensively assess the effectiveness of the Flush+Reload-based attack. The results are summarized in Table \ref{table:asrflushreload}.

The primary factors affecting the success of the Flush+Reload-based attack are false positives and false negatives. Therefore, we adopt the Recall/Precision/F1 score to evaluate the accuracy of our attack. False positives mainly arise when the hyperparameter $U$ in the DAA is set too low, causing multiple cache hits within a single function invocation to be incorrectly counted as multiple events. In contrast, false negatives are primarily caused by two factors: (1) environmental noise, which may evict the cache line before the reload phase, and (2) an overly large $U$, which may merge multiple function invocations into a single detected event. For each tool, the attacker calibrates $U$ once on the victim’s system and reuses the calibrated value for subsequent invocations. The same value can also be reused across systems that run an identical tool binary.

Overall, our results demonstrate consistently high attack effectiveness, with most configurations achieving near-1 F1 scores and the lowest performance still reaching 93.9\%. The relatively lower accuracy occurs in testing the Linux command \texttt{cd}. For \texttt{cd}, increasing $U$ leads to more false negatives, while decreasing $U$ introduces more false positives. This is because a sufficiently large number of consecutive cache misses is required to ensure that a single function invocation produces only one detected cache hit; however, when two invocations occur in close temporal proximity, they may be incorrectly merged into one. We emphasize that the invocation intervals used in our experiments are more stringent than those in real-world scenarios. In practice, consecutive invocations of the same tool are typically separated by several seconds or even tens of seconds, which would further improve attack accuracy.

\subsection{\emph{Prime}+\emph{Probe}-based Attacks}
\label{subsection:p+p}
\emph{Prime}+\emph{Probe} can also be leveraged to implement tool-usage attacks. Compared to \emph{Flush}+\emph{Reload}, it does not require the adversary to share memory with the victim, enabling stronger attack capabilities such as cross-VM attacks.

Given that modern CPUs predominantly adopt non-inclusive cache hierarchies, we target the Snoop Filter (SF) as the primary side-channel surface. To improve the robustness of the attack in noisy environments (e.g., public cloud settings), we employ \cite{zhao2024last} to identify eviction sets for the SF and to reliably perform Prime+Probe measurements. The detailed attack procedure is as follows:

\begin{itemize}[leftmargin=*]
    \item \textbf{Prime:} Prime SF sets corresponding to the victim address.
    \item \textbf{Wait:} Sleep for a predefined time window to allow the victim to execute.
    \item \textbf{Probe:} Probe eviction sets to detect the eviction in the SF caused by accesses to the victim address.
\end{itemize}

\textbf{Handle the Large Size of SF Sets.} For modern CPUs, the number of SF sets is typically large. For example, Intel Xeon Platinum 8369B features 65536 SF sets, while providing only 64 logical cores. In practice, the adversary cannot utilize all cores for the attack, as many are occupied by other users or tenants. Under this setting, it is infeasible to simultaneously monitor all 65536 SF sets. To address this issue, we use Power Spectral Density (PSD) to identify victim-related SF sets and reduce the monitoring overhead. Implementation details are provided in Appendix \ref{appedix:hlsp+p}. %In our experiments, we limit the adversary to 16 CPU cores and monitor 64 victim-related SF sets.

\textbf{Tool fingerprinting.} For \emph{Flush}+\emph{Reload} attacks, a cache hit on any address within the code segment of a tool library directly indicates that the victim LLM has invoked that tool. In contrast, for \emph{Prime}+\emph{Probe} attacks, the adversary can only observe cache set access traces and cannot directly attribute a given trace to a specific tool invocation.

\begin{figure}[tb]
	\centering
	\subfigure[Linux Command \texttt{cp}.]{\includegraphics[width=4.1cm]{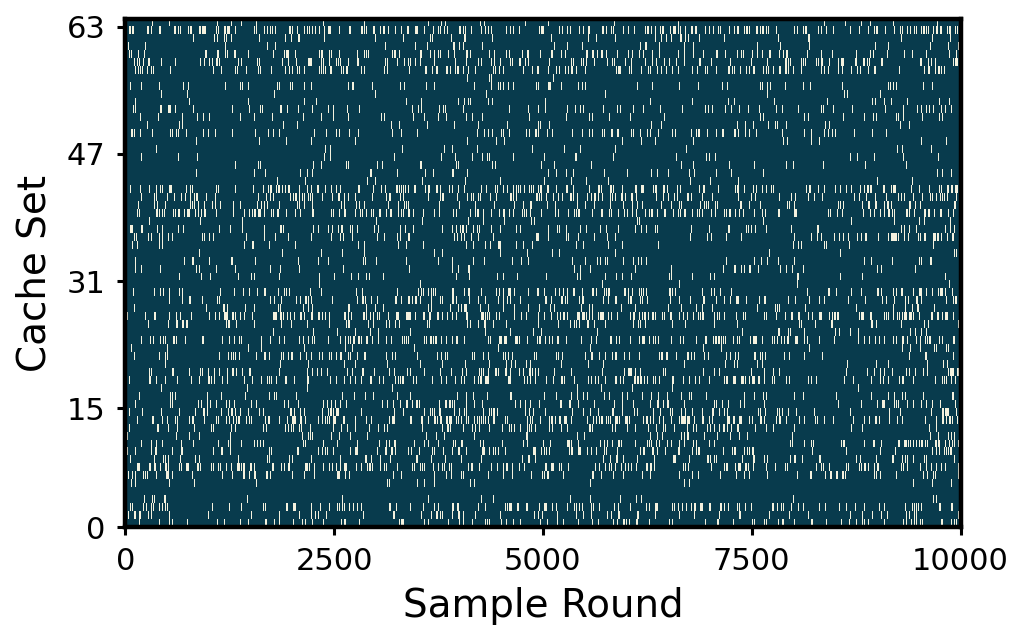}}
	\subfigure[Git Command \texttt{git commit}.]{\includegraphics[width=4.1cm]{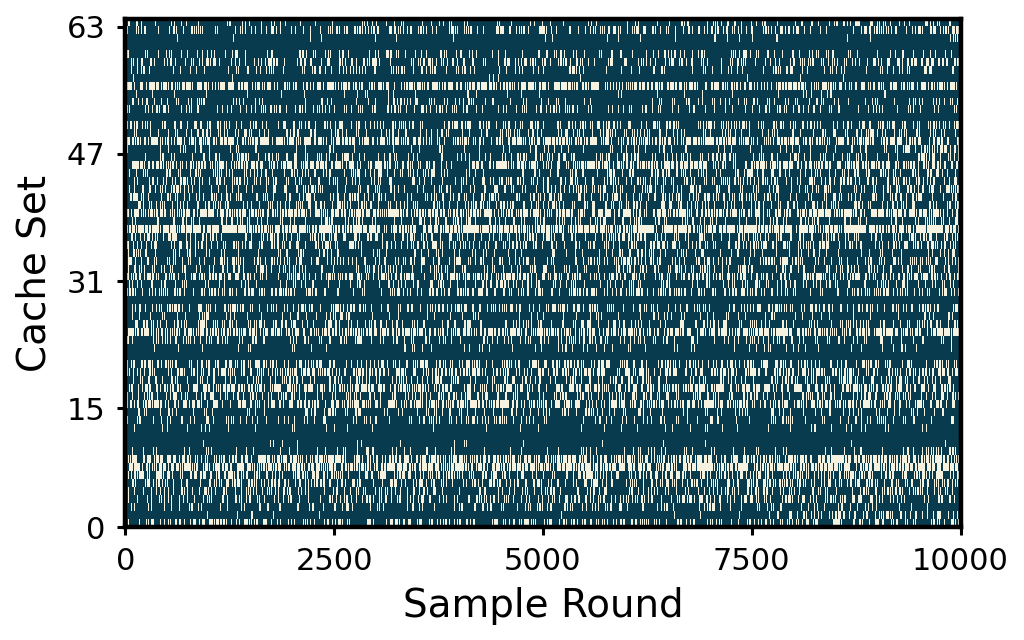}}
	\caption{\emph{Prime}+\emph{Probe} Attack Memorygrams on \texttt{cp} and \texttt{git commit}.}
    %\vspace{-0.2cm}
    \label{figure:memorygram}
\end{figure}

To enable tool fingerprinting, we first transform the collected traces into memorygrams, which capture cache hit/miss patterns within a time window during \emph{Prime}+\emph{Probe} execution. Memorygrams have been widely used in CPU/GPU side-channel attacks, where the captured access patterns are leveraged to infer victim activity. Figure \ref{figure:memorygram} illustrates the memorygrams corresponding to two LLM tool invocations (Linux Command \texttt{cp} and Git Command \texttt{git commit}). We perform 10,000 rounds of \emph{Prime}+\emph{Probe} measurements; darker regions indicate that accesses to the target cache set are detected in a given round, while lighter regions indicate no detected access. We then use these memorygrams to train classifiers for distinguishing between different tools. 

\begin{figure}[tb]
	\centerline{\includegraphics[width=8.5cm]{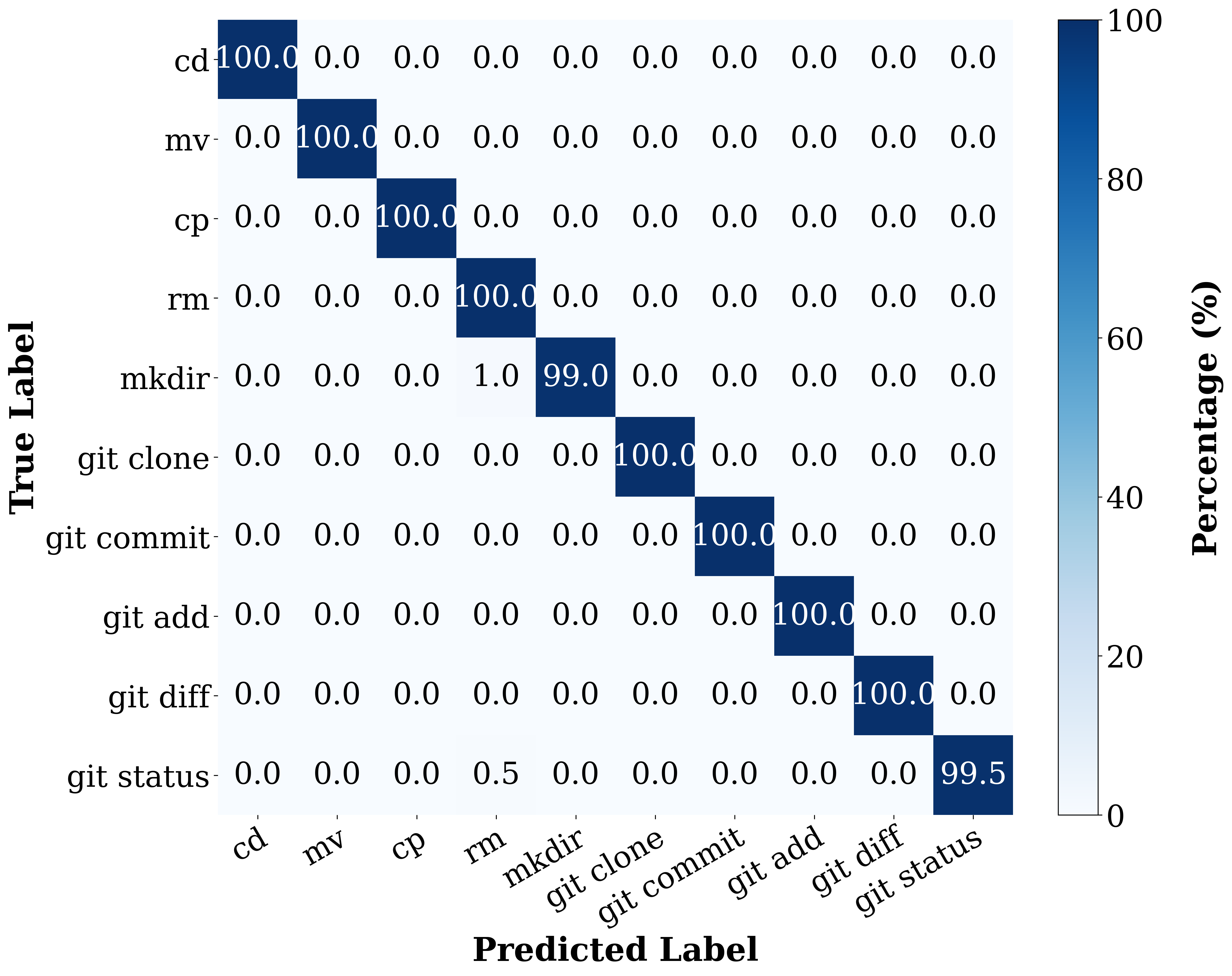}}
	\caption{Confusion Matrix of the Classifier for the 10 LLM Tools \emph{Prime}+\emph{Probe}-based Attack.}
	\label{figure:confmatrix}
\end{figure}

\textbf{Real-world evaluation.} To ensure robustness, we evaluate our \emph{Prime}+\emph{Probe} attack in public cloud environments, similar to our \emph{Flush}+\emph{Reload} experiments. We collect 1000 memorygrams for each of ten LLM tools across two categories on public cloud instances. Notably, memorygrams are collected within a fixed time window (10000 sampling rounds). Because it is difficult to align precisely with the start and end of tool execution, a given window may not capture the full access pattern of a tool invocation. Nevertheless, our subsequent experiments show that even partial access patterns are sufficient for accurate tool fingerprinting. We train an image classifier based on ResNet-18 using 700 memorygrams per tool for training, 100 for validation, and 200 for testing. The final results, shown in Figure \ref{figure:confmatrix}, demonstrate that tool fingerprinting achieves high accuracy across all evaluated tools.

\textbf{Limitation of \emph{Prime}+\emph{Probe}.} Although \emph{Prime}+\emph{Probe} enables cross-VM attacks and therefore has a stronger threat model than \emph{Flush}+\emph{Reload}, it still faces the following limitations when used to implement an end-to-end tool-workflow recovery attack:

\begin{itemize}[leftmargin=*]
    \item \textbf{Higher overhead:} \emph{Prime}+\emph{Probe} incurs higher overhead than \emph{Flush}+\emph{Reload} even when monitoring a single cache location. More importantly, identifying one tool invocation requires simultaneously monitoring 64 cache sets, whereas \emph{Flush}+\emph{Reload} only needs to monitor a single cache line. As the number of monitored tools increases, the overhead of the \emph{Prime}+\emph{Probe}-based attack grows substantially and becomes significantly higher than that of the \emph{Flush}+\emph{Reload}-based attack.
    \item \textbf{Cache-footprint aliasing:} Different tool invocations may access overlapping cache sets, causing their cache footprints to become entangled. Such footprint aliasing makes different tools more difficult to distinguish and may reduce classification accuracy, especially when multiple tools are invoked within a short time window.
    \item \textbf{Lack of real-time recovery:} \emph{Prime}+\emph{Probe} cannot directly recover a tool invocation in real time. The attacker must first collect a sufficiently long cache trace, convert it into a memorygram, and then feed it into a classifier to identify the invoked tool. However, as discussed in our subsequent threat analysis, some attacks require the adversary to recover tool invocations in real time.
\end{itemize}

Therefore, in the following security analysis, we use \emph{Flush}+\emph{Reload} to implement our end-to-end attacks.

\section{Attack Design of \alias}
In this section, we present a cache-based side-channel eavesdropping framework, \alias. We implement an end-to-end agentic \emph{Tool Recovery Attack} based on \emph{Flush}+\emph{Reload}. Through two representative attack scenarios, we further investigate what private information an adversary can infer after recovering the agent’s tool-execution trace.

\subsection{Tool Recovery Attack}
\label{subsection:toolrecover}
\begin{figure}[tb]
	\centerline{\includegraphics[width=8.5cm]{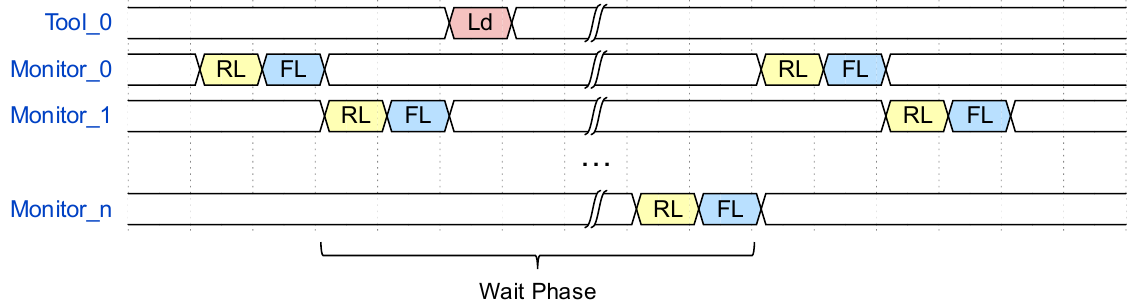}}
	\caption{Pipeline of \emph{Tool Recovery Attack}: Tool denotes the invocation status of the monitored tools, while Monitor indicates the \emph{Flush}+\emph{Reload} monitor associated with each tool. FL and RL represent the \emph{Flush} and \emph{Reload} phases, respectively.}
	\label{figure:end-to-end}
\end{figure}

In Section \ref{subsection:flush+reload}, we demonstrate that individual agentic tool invocations can be eavesdropped on using \emph{Flush}+\emph{Reload}. However, fully recovering an agent’s tool-execution trace in an end-to-end attack requires simultaneously monitoring invocations of multiple tools. To address this challenge, we implement the \emph{Tool Recovery Attack} illustrated in Figure \ref{figure:end-to-end}.

The key insight is to exploit the wait phase between the flush and reload phases. In conventional Flush+Reload attacks, the attacker typically calls \texttt{sleep()} or \texttt{sched\_yield()} during this interval to allow the victim to execute. Instead of leaving the attacker idle, our attack interleaves the flush and reload operations for other monitored tools within these wait intervals. This design enables efficient concurrent monitoring of multiple tools and supports end-to-end recovery of the agent’s tool-execution trace.

Note that the number of tools monitored concurrently cannot grow without bound. As more tools are monitored, the wait phase becomes longer, which may increase false negatives. Specifically, a cache line loaded by the victim may be evicted from the cache before the corresponding reload phase occurs, causing the attacker to miss the access. In the subsequent attack scenarios, we analyze how to minimize the number of monitored tools while preserving the effectiveness of the end-to-end attack.

\subsection{Noise Analysis and Model Input}
\alias is subject to both cache-level and tool-level noise. Cache activity from other users and prefetching can cause false positives and false negatives when detecting accesses to monitored cache lines. We evaluate these effects in detail in the public cloud experiments described in Section \ref{subsection:flush+reload}. Tool-level noise comes from invocations outside the agent’s workflow, including those triggered by background OS processes and those initiated manually by other users.

\begin{figure}[tb]
	\centerline{\includegraphics[width=8.7cm]{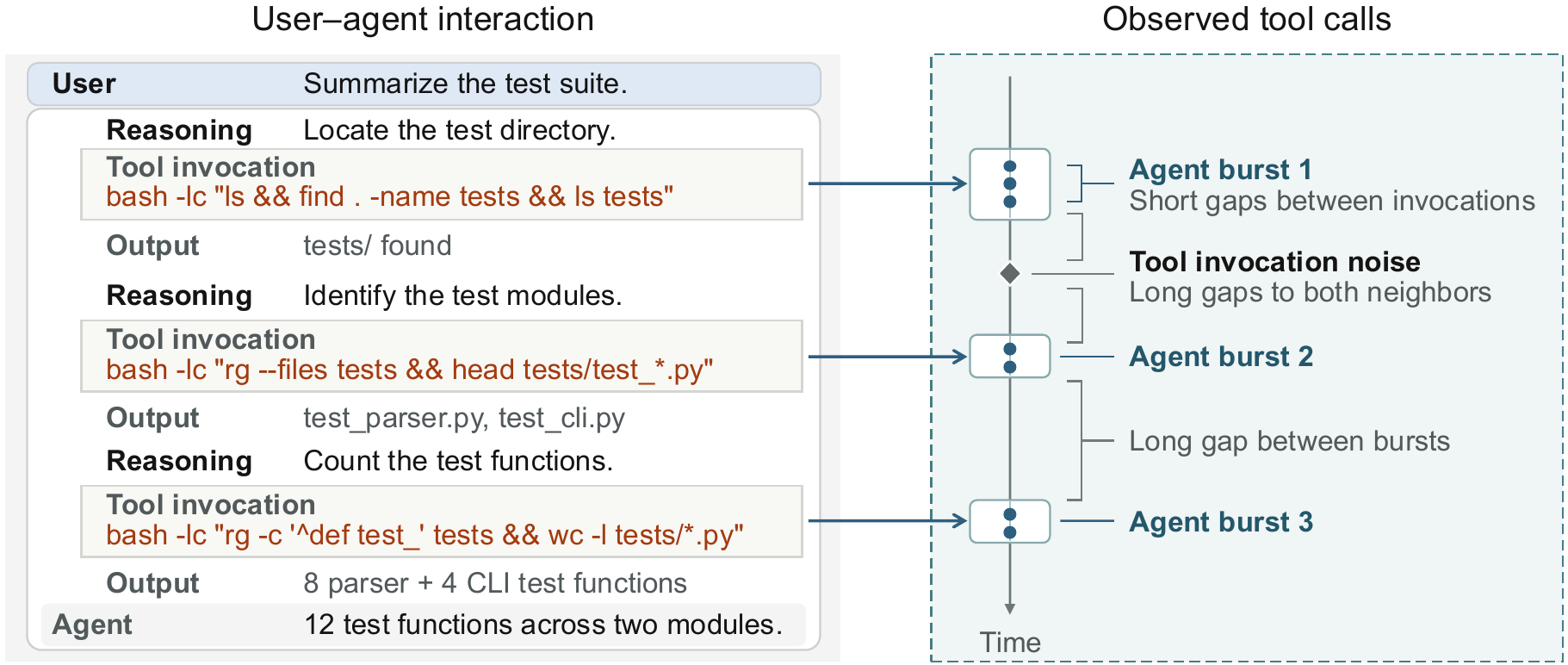}}
	\caption{Illustration of LLM Agent Tool Invocations and Their Manifestations in Observed Traces.}
	\label{figure:burst}
\end{figure}

To mitigate tool-level noise, we examine how LLM agents invoke tools. As illustrated in Figure \ref{figure:burst}, the agent issues a batch of tool invocations through \texttt{bash -lc "..."} after each reasoning step and waits for the results before resuming reasoning. Invocations within a batch tend to occur close together in time, with longer gaps between batches. Invocations outside these bursts are more likely to be tool-level noise.

We therefore represent each observed invocation by its tool identity and timestamp. The tool sequence helps the model assess whether an invocation fits the task context. For example, an unexpected \texttt{gdb} invocation in a trace of Python model training is likely to be noise. The timestamp provides an additional signal that the model can use to weigh each invocation in downstream attack tasks.

A tool that runs for an extended period can also create a long gap within a burst. For example, when \texttt{python} trains a model, the next invocation may occur late enough to appear outside the burst. We expect the model to learn to interpret such gaps in context. If the surrounding sequence suggests a model training task, a long gap after \texttt{python} is consistent with the time needed for training, and the next invocation may still be part of the agent’s workflow.

DAA consolidates multiple cache hits associated with a single tool invocation. In the attack trace, we record the timestamp of the first hit as the invocation time.
\subsection{Scenario 1: Task Profiling Attack}

\subsubsection{Adversary Goal}
As shown in Figure \ref{figure:overview}, the first attack scenario is \emph{Task Profiling Attack}, where the adversary attempts to infer the task being executed by the agent from the recovered tool workflow. Formally, a recovered trace is represented as
\begin{equation}
    \mathcal{T}_i =
    \left((t_{i,1},a_{i,1}),\ldots,(t_{i,L_i},a_{i,L_i})\right),
\end{equation}
where $t_{i,j}$ denotes the tool invoked at position $j$, $a_{i,j}$ is its recovered invocation timestamp, and $L_i$ is the trace length. Each trace is associated with a task label $y_i \in \{1,\ldots,C\}$, where $C$ denotes the number of task categories. We follow the taxonomy used in Terminal-Bench \cite{merrill2026terminal}. Table \ref{table:taxonomy} in Appendix \ref{appendix:taxonomy} summarizes all task categories. The adversary's goal is to estimate
\begin{equation}
    \hat{y}_i =
    \mathop{\arg\max}\limits_{c \in \{1,\ldots,C\}}
    P\left(y_i=c \mid \mathcal{T}_i\right).
\end{equation}

\subsubsection{Attack Modeling}
\label{subsubsection:taskprofilingmodel}
Given a recovered tool trace, the adversary trains a classifier to infer the underlying task category. The model jointly encodes tool semantics, invocation order, and call timing. Detailed methodology with equations is deferred to Appendix~\ref{appendix:taskprofiling}.

\textbf{Semantic Tool Representation.} Instead of assigning independent embeddings to tools, which ignores functional similarity among tools, we initialize tool representations using a pretrained sentence encoder \cite{awasthy2026graniteembeddingmultilingualr2}. Specifically, each tool name is mapped into a semantic embedding space, allowing tools with similar functions (e.g., file search and file reading) to have related representations. These embeddings provide the tool-identity features for trace encoding.

\textbf{Bidirectional Trace Encoding.} Tool traces exhibit sequential dependencies, where different tasks may invoke similar tools but follow different execution orders. For each invocation, we also derive the elapsed time since the first recorded call and the gap since the preceding call. Both timing features are expressed in milliseconds, transformed with $\log(1+x)$, and standardized using statistics fitted only on training traces. A two-layer MLP maps the normalized features to a time embedding. We concatenate this time embedding with the semantic tool embedding to obtain the event vector $\mathbf{v}_{i,j}$. The resulting input sequence is
\begin{equation}
    \mathbf{X}_i =
    (\mathbf{v}_{i,1},\ldots,\mathbf{v}_{i,L_i}).
\end{equation}
We then employ a bidirectional gated recurrent unit (BiGRU) to encode the trace. The forward and backward GRUs process $\mathbf{X}_i$ in opposite directions, capturing the preceding and subsequent contexts of each invocation. We concatenate the final hidden states from both directions as the trace representation:
\begin{equation}
    \mathbf{z}_i =
    \overrightarrow{\mathbf{h}}_{i,L_i}
    \mathbin{|}
    \overleftarrow{\mathbf{h}}_{i,1}
    \in \mathbb{R}^{2d_h},
\end{equation}
where $\overrightarrow{\mathbf{h}}_{i,L_i}$ and $\overleftarrow{\mathbf{h}}_{i,1}$ denote the final hidden states of the forward and backward GRUs, respectively, and $d_h$ is the hidden dimension of each GRU direction. Since the attack is performed after collecting the complete trace, the adversary can use bidirectional context during inference.

\textbf{Task Classification.} The trace representation is fed into a multilayer perceptron classifier that outputs the probability distribution over task categories. The model is trained with categorical cross-entropy using labeled traces. The classifier jointly models tool semantics, invocation order, and call timing to infer the task category.

%\subsubsection{Dataset}

% \subsubsection{Impact of Varying the Number of Monitored Tools}
% \shinan{why here? could be in the evaluation 6.3}

\subsection{Scenario 2: Trigger Localization Attack}
\subsubsection{Adversary Goal}
The second attack scenario is \emph{Trigger Localization Attack}. As illustrated in Figure \ref{figure:overview}, the adversary aims to identify sensitive tools such as cryptographic operations or model inference, thereby enabling more fine-grained attacks. For example, the attacker may subsequently launch specialized side-channel attacks to extract cryptographic keys \cite{yan2019attack,yarom2014flushreload,liu2015last}, recover model parameters \cite{yan2020cache,zhang2023t}, or reconstruct model outputs \cite{ding2025moecho,gao2025iknow}. Although such fine-grained attacks can reveal substantially more sensitive information, they are typically much more costly to execute continuously. \emph{Trigger Localization Attack} therefore serves as a lightweight trigger that identifies when the target sensitive operation occurs, allowing the adversary to activate expensive fine-grained attacks only when necessary.

Let a recovered tool-execution trace be represented as
\begin{equation}
    \mathcal{T}
    =
    \left((x_1,a_1),(x_2,a_2),\ldots,(x_L,a_L)\right),
\end{equation}
where $x_t$ denotes the tool invoked at position $t$, $a_t$ is its recovered invocation timestamp, and $L$ is the trace length. Each invocation has a binary state label $y_t \in \{0,1\}$: $y_t=1$ identifies a sensitive invocation, while $y_t=0$ identifies a non-sensitive invocation. These labels provide supervision during training and are unavailable at inference time.

For every position $t$, the adversary estimates
\begin{equation}
    P\left(y_t \mid x_{1:t},a_{1:t}\right).
\end{equation}
The prediction at position $t$ uses only the observed tools and invocation timestamps through that position. Future invocations and their timestamps are unavailable to the model. This restriction preserves causal inference during validation and deployment and supports online localization as the victim's workflow unfolds.

Our localization objective is inherently asymmetric. A false negative causes a sensitive tool invocation to be missed, potentially preventing the attacker from capturing the corresponding sensitive data. In contrast, a false positive only causes an additional tool invocation to be included in subsequent monitoring, resulting in extra overhead without directly compromising attack completeness. We therefore treat false negatives as substantially more costly than false positives. However, labeling every invocation as sensitive trivially achieves zero false negatives while incurring maximum overhead. Our objective is thus to minimize the fraction of invocations labeled as sensitive subject to a predefined recall constraint:
\begin{equation}
    \min_{\tau}\;O_m(\tau)
    \quad
    \mathrm{s.t.}\quad
    \mathrm{R}(\tau)\geq R_{\min}.
    \label{eq:trigger_objective}
\end{equation}
Here, $\tau$ is the decision threshold, $\mathrm{R}$ is sensitive-invocation recall, and $O_m$ is the proportion of invocations labeled as sensitive. We use $O_m$ as a proxy for the monitoring overhead of a subsequent attack.

\subsubsection{Attack Modeling}
\label{subsubsection:trigger_attack_model}
Since trigger localization requires online identification of sensitive tool invocations instead of offline trace-level classification, we design a sequence model tailored for this setting (with details deferred to Appendix~\ref{appendix:triggerLocalization}).

\textbf{Unidirectional Sequence Labeler.} We use the semantic tool representation described in Section~\ref{subsubsection:taskprofilingmodel} and the timing-feature construction used for task profiling. At position $t$, we concatenate the semantic embedding $\mathbf{e}_{x_t}$ with an MLP encoding of elapsed time and the inter-call gap to form $\mathbf{v}_t$. Both timing features depend only on invocation timestamps observed through position $t$. A unidirectional GRU updates its hidden state as
\begin{equation}
    \mathbf{h}_t =
    \operatorname{GRU}
    (\mathbf{v}_t,\mathbf{h}_{t-1}).
\end{equation}
The hidden state $\mathbf{h}_t$ summarizes the current invocation and its observed history, including tool identities and timing. A lightweight classification head maps $\mathbf{h}_t$ to the probability that the current invocation is sensitive. The model can update this prediction incrementally as each invocation is observed.

\textbf{Class-Imbalance-Aware Training.}
Sensitive tool invocations are typically rare, and false negatives are more costly because they may cause the attacker to miss critical operations. We therefore train the sequence labeler using a class-weighted cross-entropy loss, which assigns a larger penalty to errors on sensitive invocations. This prevents the model from achieving high accuracy by simply labeling most tool invocations as non-sensitive, while reducing the monitoring overhead under a low false-negative requirement.

\textbf{Recall-Constrained Model Selection.}
As shown in Equation 7, our goal is to reduce the attacker’s monitoring overhead as much as possible while satisfying a predefined minimum recall requirement. To achieve this goal, we perform recall-constrained model selection. For each checkpoint obtained at epoch $e$, we adjust its decision threshold $\tau$ to satisfy a predefined minimum recall $R_{\min}$ and select the threshold that  minimizes $O_{m}$. After selecting the model, we discard its model-validation threshold and use an independent calibration set to determine the final threshold under the same recall-constrained objective. The resulting model and threshold are fixed before test-time evaluation.
\section{Evaluation}
In this section, we evaluate the end-to-end \emph{Tool Recovery Attack} and assess its effectiveness in the two attack scenarios.

\begin{table}[t]
	\caption{Evaluation of \emph{Tool Recovery Attack}.} 
    \small
	\begin{center}
		\setlength\tabcolsep{3pt}
		\begin{threeparttable}
			\begin{tabular}{c|c|c|c|c}
				\toprule[2pt]
                Agent & Model & Recall & Precision & F1\\
                \hline
                \multirow{5}{*}{Codex} & gpt-6.1-sol & 95.2\% & 97.5\% & 96.3\%\\
                \cline{2-5}
                 & gpt-6-astra & 91.3\% & 94.9\% & 93.1\%\\
                \cline{2-5}
                 & gpt-5.6-sol & 100\% & 97.6\% & 98.8\% \\
                \cline{2-5}
                 & gpt-5.6-terra &  98.1\% & 96.3\% & 97.2\%\\
                \cline{2-5}
                 & gpt-5.6-luna & 94.6\% & 98.1\% & 96.3\%\\
                % \cline{2-5}
                % & glm-5.3 & 99.1\% & 97.1\% & 98.1\%\\
                %\cline{2-5}
                % & glm-4.7 & 96.7\% & 90.7\% & 93.6\%\\
                \hline
                \multirow{6}{*}{Claude Code} & claude-fable-5 &  96.8\% & 100\% & 98.4\%\\
                \cline{2-5}
                 & claude-opus-5 & 99.1\% & 96.9\% & 98.0\%\\
                \cline{2-5}
                 & claude-sonnet-5 & 95.6\% & 97.3\% & 96.4\%\\
                \cline{2-5}
                 & claude-haiku-4.5 & 91.2\% & 96.9\% & 93.9\%\\
                \cline{2-5}
                 & glm-5.3 & 92.8\% & 98.5\% & 95.6\%\\
                \cline{2-5}
                 & glm-4.7 & 98.3\% & 96.7\% & 97.5\%\\
				\bottomrule[2pt]
			\end{tabular}
		\end{threeparttable}
		\label{table:evaluationtra}
	\end{center}
    \vspace{-0.5cm}
\end{table}

\begin{figure*}[t]
	\centering
	\subfigure[All Tools.]{\includegraphics[width=4.3cm]{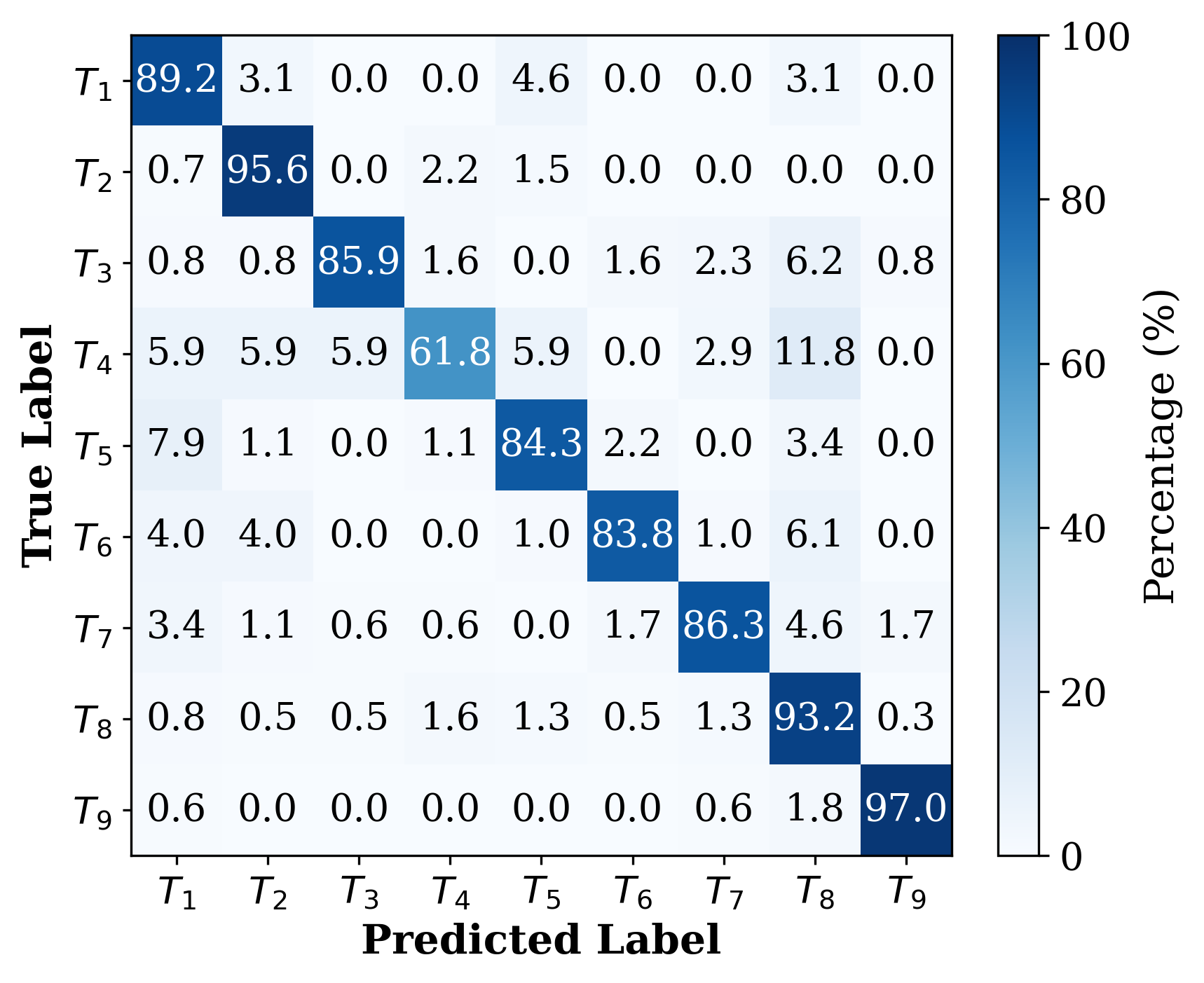}\label{figure:alltools}}
	\subfigure[Top-100 Tools.]{\includegraphics[width=4.3cm]{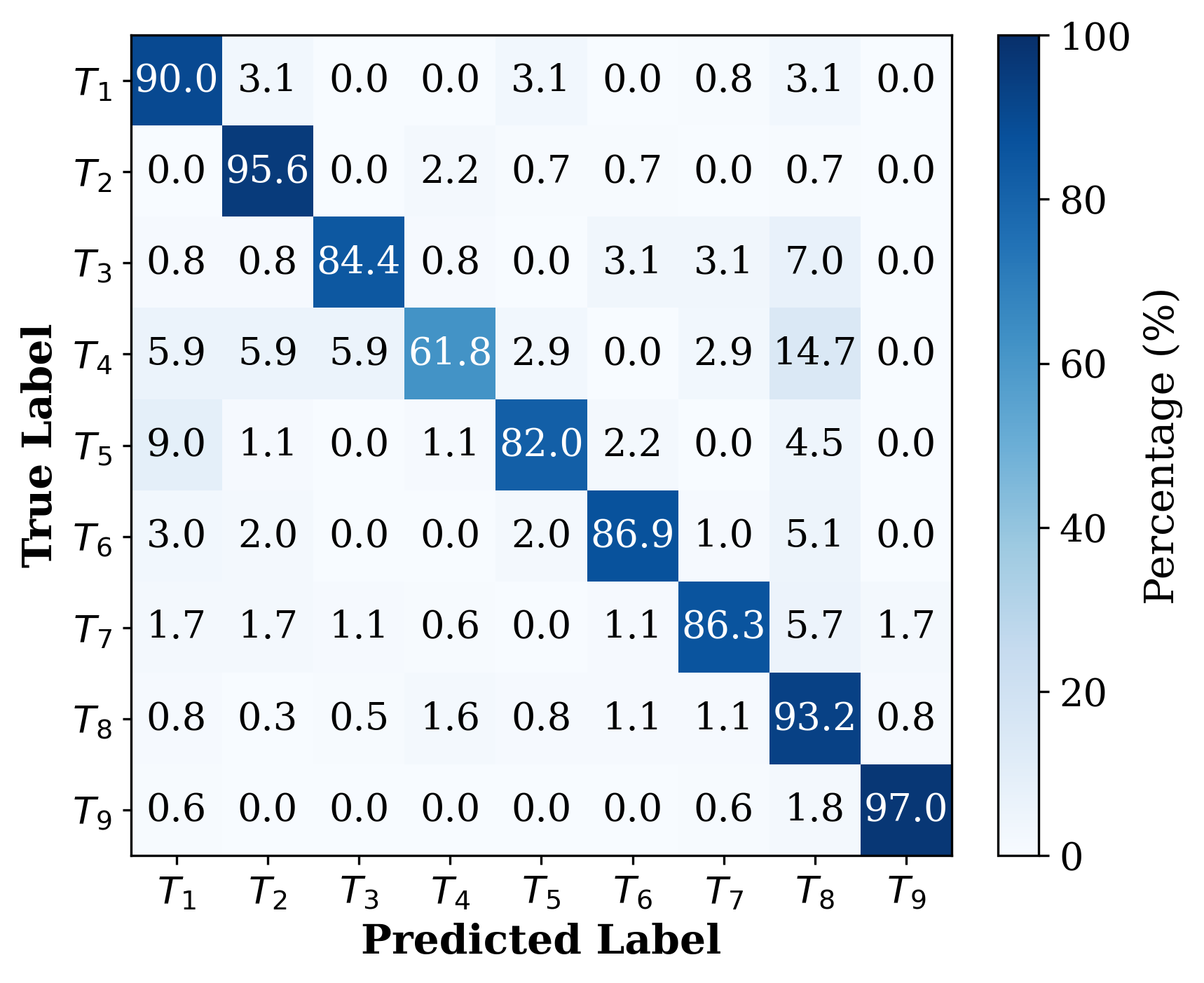}\label{figure:top100}}
    \subfigure[Top-50 Tools.]{\includegraphics[width=4.3cm]{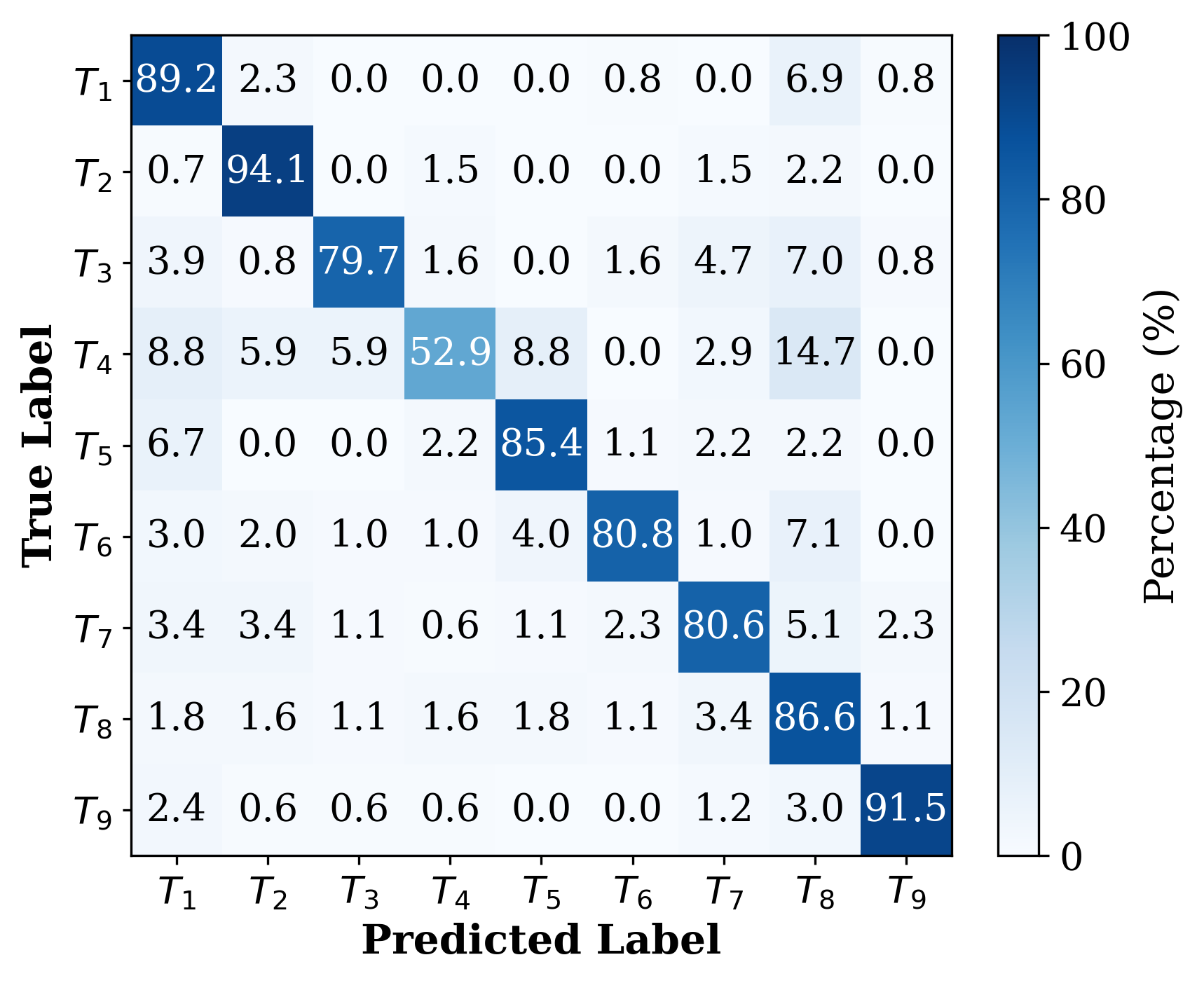}\label{figure:top50}}
    \subfigure[Top-20 Tools.]{\includegraphics[width=4.3cm]{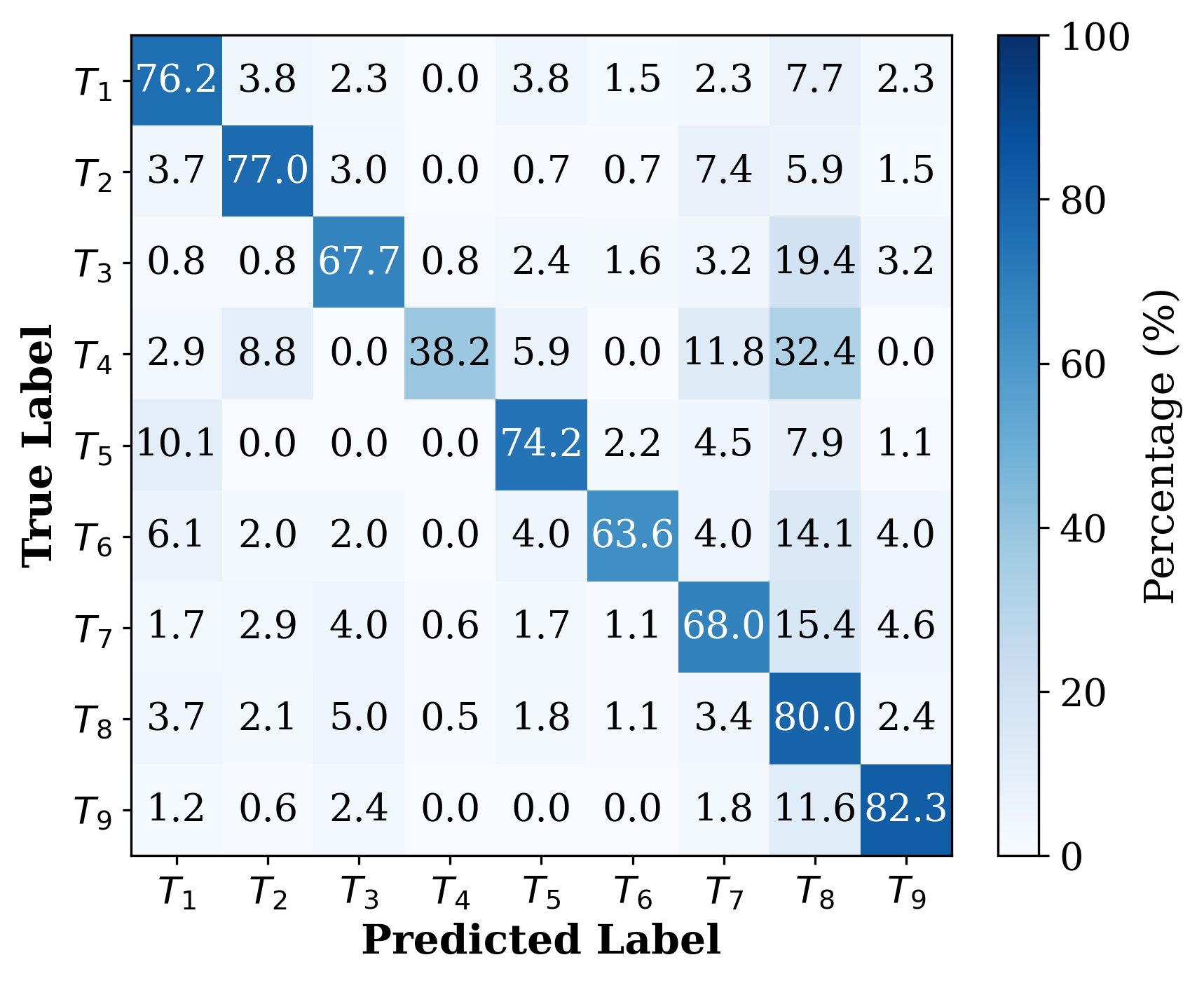}\label{figure:top20}}
	\caption{Confusion Matrices for \emph{Task Profiling Attack} Using Traces Obtained by Monitoring All, the Top-100, the Top-50, and the Top-20 Tools.}
    \label{figure:toptools}
\end{figure*}

\subsection{Evaluation of Tool Recovery Attack}
We first evaluate whether the \emph{Tool Recovery Attack}, which simultaneously monitors 100 tools and reconstructs the victim agent’s tool-invocation trace, suffers any performance degradation compared with the single-tool \emph{Flush}+\emph{Reload} attack evaluated in Section \ref{subsection:flush+reload}. The rationale for monitoring 100 tools is discussed in Section \ref{subsubsection:sentools}. We evaluate the \emph{Tool Recovery Attack} by running Terminal-Bench and our custom tasks designed for \emph{Trigger Localization Attack} across 11 models in Codex and Claude Code. Meanwhile, we collect the recovered tool-invocation traces to construct the datasets used to train the models for the two attack scenarios.

We conduct all evaluation on a real public-cloud server instance equipped with an Intel Xeon Platinum 8163 CPU and an NVIDIA T4 GPU. Our measurements span all periods of the day to capture varying levels of background activity and evaluate the robustness of the \emph{Tool Recovery Attack} under noisy real-world cloud conditions.

%\shinan{what is the workload?}
%We evaluate the Tool Recovery Attack on 11 models across two agents, Codex and Claude Code. 
During all experiments, we enable the sandbox provided by each agent. As shown in Table \ref{table:evaluationtra}, although the F1 scores exhibit a slight decrease compared with the single-tool \emph{Flush}+\emph{Reload} attack, the \emph{Tool Recovery Attack} still maintains consistently high performance across all evaluated models, with the lowest F1 score remaining above 93\%.

\subsection{Evaluation of Task Profiling Attack}
\label{subsubsection:sentools}
\textbf{Dataset.} In the \emph{Task Profiling Attack}, we run Terminal-Bench 1 and Terminal-Bench 2 \cite{merrill2026terminal} across 11 models in Codex and Claude Code, collecting 4000 tool-invocation traces to construct the dataset for evaluating the attack.

\textbf{Sensitivity to the Number of Monitored Tools.} As discussed in Section \ref{subsection:toolrecover}, the number of tools that can be monitored concurrently is limited because monitoring more tools incurs higher overhead. In this section, we evaluate how reducing the number of concurrently monitored tools affects \emph{Task Profiling Attack} accuracy.

We enumerate all tools appearing in the ground-truth traces of our dataset, yielding a total of 378 tools. We first count the frequency of each of the 378 tools across all traces in the dataset and retain the most frequently invoked tools, with the goal of reducing the number of monitored tools without sacrificing classification accuracy. We evaluate the model under four settings: using the complete tool traces, and retaining only the top-100, top-50, and top-20 most frequently invoked tools. All other experimental settings, including the training configuration and train/test split, are kept unchanged.

Figure \ref{figure:toptools} shows the results. The classification accuracy is 89.81\% when using all tools, 89.81\% with the top-100 tools, 85.38\% with the top-50 tools, and 74.21\% with the top-20 tools. Monitoring only the top-100 tools achieves nearly identical accuracy to monitoring all tools, whereas reducing the monitored set to the top-50 tools leads to an approximately 4\% accuracy drop, and the degradation becomes much more pronounced with only the top-20 tools. Therefore, in our end-to-end attack, we monitor the selected top-100 tools to substantially reduce monitoring overhead while preserving task-classification accuracy.

\textbf{Attack Performance.} We further evaluate the \emph{Task Profiling Attack} using the traces collected from our end-to-end attack that monitors the top-100 tools. As shown in Table \ref{table:tpa}, the attack achieves accuracies of 87.77\% and 91.39\% on Terminal-Bench 1 and Terminal-Bench 2, respectively. We also report the macro-F1 and weighted-F1 scores across all task categories. For both datasets, we observe that macro-F1 is lower than weighted-F1 for most models. This is mainly because the Games category contains substantially fewer tasks than the other categories in Terminal-Bench, resulting in fewer corresponding traces in our dataset and consequently lower classification performance for this class. Since macro-F1 assigns equal weight to every category, the poorer performance on Games has a larger impact on the overall score. In contrast, weighted-F1 accounts for class frequency and is therefore less affected by this minority category. We expect this limitation to be mitigated by collecting additional training traces for the Games category.

\begin{table*}[t]
	\caption{Evaluation of \emph{Task Profiling Attack}.} 
    \footnotesize
	\begin{center}
		\setlength\tabcolsep{3.8pt}
		\begin{threeparttable}
			\begin{tabular}{c|c|c|c|c|c|c|c|c|c|c|c|c|c}
				\toprule[2pt]
                \multicolumn{2}{c|}{Agent} & \multicolumn{5}{c|}{Codex} & \multicolumn{6}{c|}{Claude Code} & \multirow{2}{*}{Overall}\\
                \cline{1-13}
                \multicolumn{2}{c|}{Model} & \makecell{gpt-\\6.1-sol} & \makecell{gpt-\\6-astra} & \makecell{gpt-\\5.6-sol} & \makecell{gpt-\\5.6-terra} & \makecell{gpt-\\5.6-luna} & \makecell{claude-\\fable-5} & \makecell{claude-\\opus-5} & \makecell{claude-\\sonnet-5} & \makecell{claude-\\haiku-4.5} & glm-5.3 & glm-4.7 & \\
                \hline
                \multirow{3}{*}{\makecell{Terminal\\Bench 1}} & Accuracy & 94.12\% & 83.33\% & 87.50\% & 77.78\% & 86.21\% & 85.71\% & 91.30\% & 94.74\% & 93.33\% & 90.91\% & 90.24\% & 87.77\%\\
                \cline{2-14}
                & Macro-F1 & 94.76\% & 76.70\% & 90.82\% & 71.69\% & 73.18\% & 79.20\% & 83.61\% & 85.71\% & 80.61\% & 91.54\% & 91.18\% & 81.98\%\\
                \cline{2-14}
                & Weighted-F1 & 93.34\% & 81.89\% & 89.10\% & 78.17\% & 85.42\% & 88.87\% & 91.11\% & 94.74\% & 91.96\% & 92.00\% & 90.78\% & 87.83\%\\
                \hline
                \multirow{3}{*}{\makecell{Terminal\\Bench 2}} & Accuracy & 86.79\% & 86.27\% & 96.23\% & 86.36\% & 92.73\% & 88.24\% & 97.06\% & 95.12\% & 88.24\% & 95.00\% & 94.59\% & 91.39\%\\
                \cline{2-14}
                & Macro-F1 & 83.21\% & 78.61\% & 96.00\% & 82.47\% & 92.10\% & 89.47\% & 94.07\% & 92.99\% & 89.47\% & 96.48\% & 84.07\% & 88.89\%\\
                \cline{2-14}
                & Weighted-F1  & 86.63\% & 87.85\% & 96.11\% & 86.22\% & 92.89\% & 88.22\% & 97.25\% & 95.35\% & 88.22\% & 94.54\% & 93.58\% & 91.40\%\\
				\bottomrule[2pt]
			\end{tabular}
		\end{threeparttable}
		\label{table:tpa}
	\end{center}
    \vspace{-0.5cm}
\end{table*}

\subsection{Evaluation of Trigger Localization Attack}
In the \emph{Trigger Localization Attack}, we consider two case studies. The first is key extraction, where the adversary aims to launch a more fine-grained side-channel attack during the victim agent’s encryption or decryption operations to recover cryptographic keys \cite{liu2015last,zhao2024last,chen2024prefetchx,jiang2016complete}. In this case, Trigger Localization aims to identify tool invocations that are likely associated with encryption or decryption. The second is model extraction, where the adversary seeks to recover parameters of models used by the victim agent through other side channels, such as GPU-based attacks \cite{dutta2023spy,zhang2023t,wei2020leaky}. Here, Trigger Localization aims to identify tool invocations associated with model training or inference. This could happen when a victim is using agents to develop better modeling for AIOps use cases \cite{chan2025mle,tang2026ai,chen2025aiopslab,jha2025itbench}.

\textbf{Dataset.} For each case study, we design 100 tasks, with representative examples provided in Appendix \ref{appendix:trigger_examples}. We then collect 3000 traces for each case study across 11 models in Codex and Claude Code to train the localization model and evaluate attack performance. During trace collection, we observe that the key-extraction tasks involve only 79 distinct tools, while the model-extraction tasks involve 49 distinct tools. Since both numbers are below 100, the corresponding end-to-end attacks monitor only these 79 and 49 tools.

\textbf{Impact of $R_{min}$ on Model Performance.} We first evaluate the impact of different $R_{min}$ values on model performance to determine the $R_{min}$ used in the \emph{Trigger Localization Attack}. We evaluate both case studies under four $R_{min}$ settings: 90\%, 95\%, 97\%, and 99\%, to study how the minimum-recall requirement affects the achieved recall and the attacker overhead (i.e. $O_{m}$). As described in Section \ref{subsubsection:trigger_attack_model}, we split the dataset into training, validation, calibration, and test sets. For each epoch, we use the validation set to select the model checkpoint and then determine its decision threshold on the calibration set.

\begin{figure}[t]
	%\vspace{-1em}
	\centering
	\subfigure[Impact on Recall.]{\includegraphics[width=4.1cm]{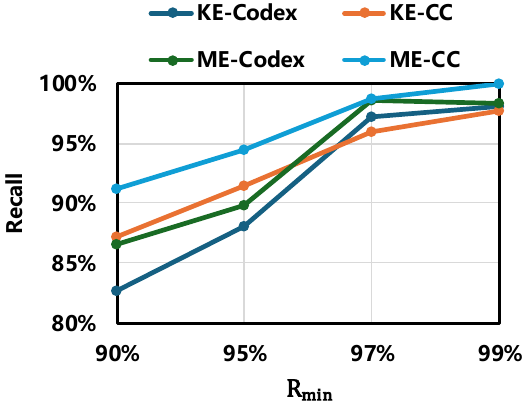}}
	\subfigure[Impact on $O_{m}$.]{\includegraphics[width=4.1cm]{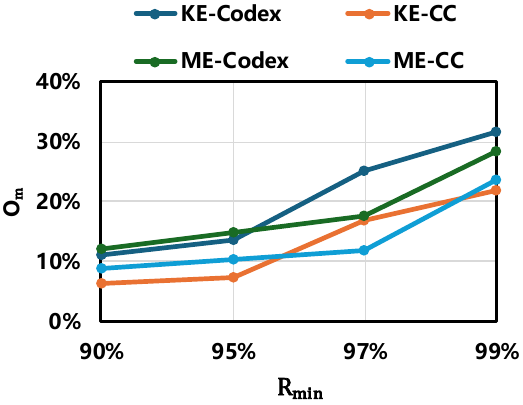}}
	\caption{Impact of $R_{min}$ on Model Performance: In the legend, KE denotes the Key Extraction case study, ME denotes the Model Extraction case study, and CC denotes Claude Code.}
    \label{figure:rmin}
    %\vspace{-0.5cm}
\end{figure}

Figure \ref{figure:rmin} shows the results. For both key extraction and model extraction, increasing $R_{min}$ leads to higher recall on the test set, but also increases $O_{m}$, revealing a clear trade-off between localization recall and attack overhead. Increasing $R_m$ from 97\% to 99\% yields only a marginal gain in test-set recall while substantially increasing $O_m$. We therefore use the model selected with $R_m = 97\%$ in each case.%For key extraction, increasing $R_{min}$ from 97\% to 99\% further improves test-set recall while introducing only a moderate increase in $O_{m}$. We therefore select the model trained with $R_{min}=99\%$. In contrast, for model extraction, increasing $R_{min}$ from 97\% to 99\% yields only a marginal recall improvement while causing a substantial increase in $O_{m}$. We therefore select $R_{min}=97\%$ for this case study.

\textbf{Attack Performance.} We then evaluate the \emph{Trigger Localization Attack} for both case studies across 11 models in Codex and Claude Code. The results are shown in Table \ref{table:tla}. For key extraction, even the worst-performing model misses only 7.04\% of sensitive tool invocations, while the model with the highest overhead requires only 39.44\% of the original monitoring overhead. Overall, the attack achieves a recall of 96.49\% while reducing the monitoring overhead to 19.54\% of the original level. For model extraction, the worst-performing model misses only 3.42\% of sensitive invocations. Overall, the attack achieves a recall of 98.61\% while reducing the overhead to 14.21\%.
\begin{table*}[t]
	\caption{Evaluation of \emph{Trigger Localization Attack}.} 
    \footnotesize
	\begin{center}
		\setlength\tabcolsep{3.8pt}
		\begin{threeparttable}
			\begin{tabular}{c|c|c|c|c|c|c|c|c|c|c|c|c|c}
				\toprule[2pt]
                \multicolumn{2}{c|}{Agent} & \multicolumn{5}{c|}{Codex} & \multicolumn{6}{c|}{Claude Code} & \multirow{2}{*}{Overall}\\
                \cline{1-13}
                \multicolumn{2}{c|}{Model} & \makecell{gpt-\\6.1-sol} & \makecell{gpt-\\6-astra} &\makecell{gpt-\\5.6-sol} & \makecell{gpt-\\5.6-terra} & \makecell{gpt-\\5.6-luna} & \makecell{claude-\\fable-5} & \makecell{claude-\\opus-5} & \makecell{claude-\\sonnet-5} & \makecell{claude-\\haiku-4.5} & glm-5.3 & glm-4.7 & \\
                \hline
                \multirow{2}{*}{\makecell{Key\\Extraction}} & Recall & 98.53\% & 100\% & 100\% & 95.05\% & 92.96\% & 100\% & 96.05\% & 98.73\% & 94.59\% & 94.12\% & 93.33\% & 96.49\%\\
                \cline{2-14}
                & $O_{m}$ & 33.20\% & 39.44\% & 18.74\% & 27.75\% & 19.88\% & 25.78\% & 10.78\% & 22.3\% & 25.81\% & 14.92\% & 15.20\% & 19.54\%\\
                \hline
                \multirow{2}{*}{\makecell{Model\\Extraction}} & Recall & 100\% & 100\% & 100\% & 96.59\% & 96.88\% & 100\% & 98.28\% & 100\% & 100\% & 98.97\% & 96.58\% & 98.61\%\\
                \cline{2-14}
                & $O_{m}$ & 21.63\% & 28.44\% & 20.70\% & 16.62\% & 12.30\% & 16.67\% & 8.24\% & 11.92\% & 18.6\% & 13.39\% & 10.02\% & 14.21\%\\
				\bottomrule[2pt]
			\end{tabular}
		\end{threeparttable}
		\label{table:tla}
	\end{center}
    \vspace{-0.5cm}
\end{table*}

\subsection{Evaluation of Multi-User Noises}
Our earlier evaluations used traces collected on public cloud instances and thus captured the effects of both cache-level noise and tool-level noise from background OS activity. In this section, we evaluate how manually introduced tool-level noise affects \alias.

We evaluate \alias under tool-level noise recorded on a server shared by approximately 20 users. This setting has substantially more concurrent user activity than an attack on a personal computer, where a malicious process deployed by the adversary would typically run on a machine used by the victim alone.

For ethical reasons, we used the shared server solely for data collection. We collected anonymized tool invocations across all processes over a 24-hour period that included peak usage hours. We then replayed the recorded invocations in a fully controlled environment, where we evaluated \emph{Task Profiling Attack} and \emph{Trigger Localization Attack}.

\begin{figure}[t]
	%\vspace{-1em}
	\centering
	\subfigure[\emph{Task Profiling Attack}.]{\includegraphics[width=4cm]{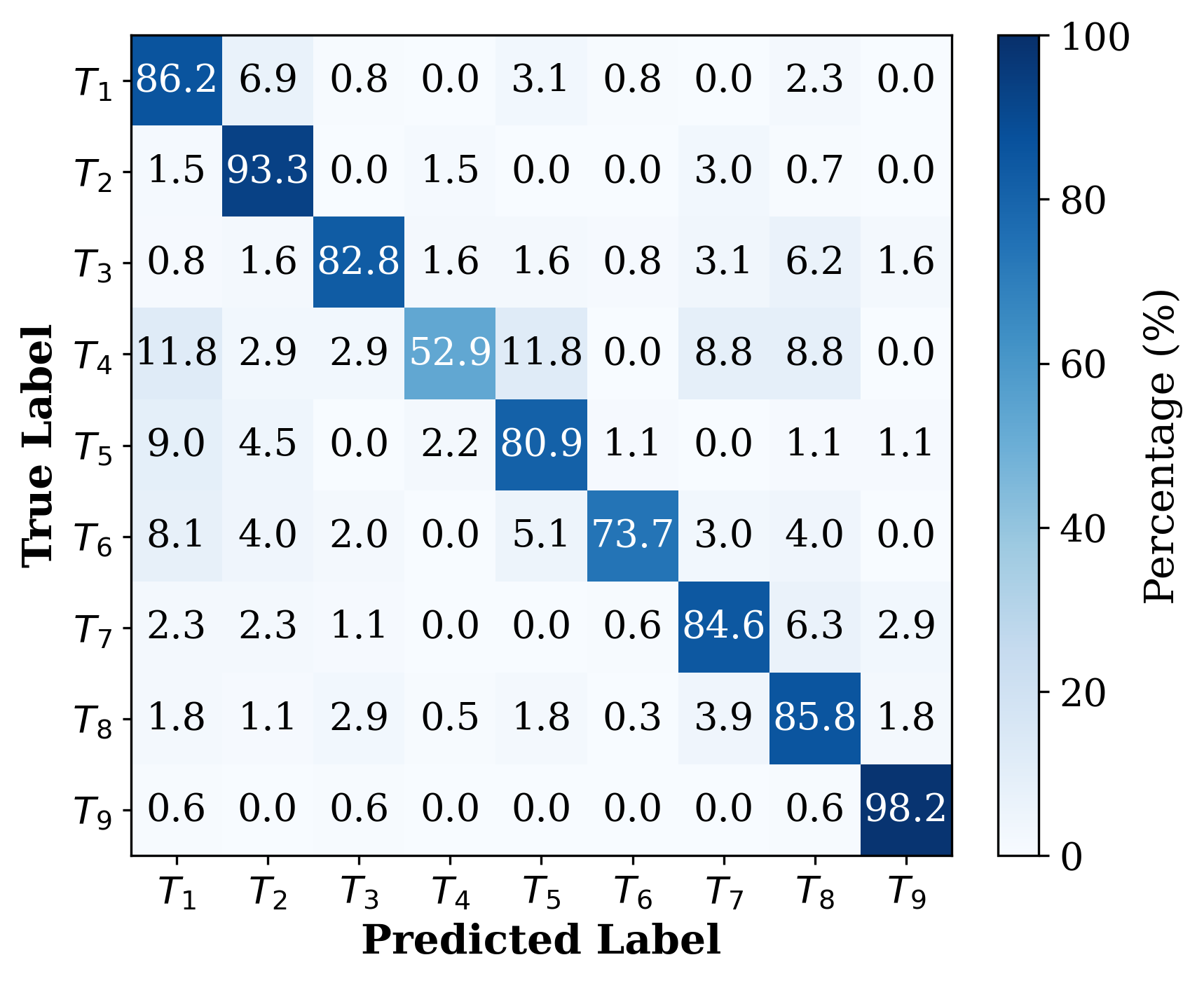}\label{subfigure:tpanoise}}
	\subfigure[\emph{Trigger Localization Attack}.]{\includegraphics[width=4.3cm]{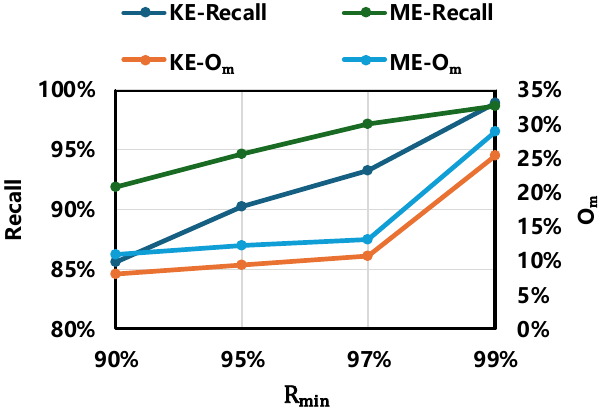}\label{subfigure:tlanoise}}
	\caption{Attack Performance under Substantial Tool-Level Noise: (a) confusion matrix of \emph{Task Profiling Attack}, (b) \emph{Trigger Localization Attack} recall and $O_m$ for key and model extraction at four $R_m$.}
    %\vspace{-0.5cm}
\end{figure}

Figure \ref{subfigure:tpanoise} shows the performance of \emph{Task Profiling Attack} under substantial tool-level noise. Across 11 models and the tasks in TB1 and TB2, \emph{Task Profiling Attack} achieves a mean accuracy of 85.61\%, approximately four percentage points below the baseline with only background tool-level noise. Figure \ref{subfigure:tlanoise} reports the performance of \emph{Trigger Localization Attack} at four values of $R_m$ under substantial noise from manually initiated tool invocations. At $R_m = 97\%$, \emph{Trigger Localization Attack} achieves a recall of 93.33\% for key extraction and 97.22\% for model extraction. At the same $R_m$, the recall of key extraction and model extraction decreases by 3\% and 1\% compared to baseline without added noise. These results show that \alias remains effective under substantial tool-level noise.
\section{Potential Countermeasures}
Potential countermeasures can be divided into two categories: eliminating cross-domain page sharing and enforcing hardware-level cache isolation. 

A direct countermeasure is to prevent mutually distrusting security domains from sharing physical pages. For example, the system can disable cross-domain page deduplication and provide each user or agent VM with private copies of tool executables and their dependent libraries. However, disabling memory deduplication mechanisms such as Kernel Same-page Merging (KSM) \cite{ksm} alone is insufficient in our setting. Processes executing the same file-backed executable or shared library may still share physical code pages through the operating system's page cache. Therefore, an effective defense must also isolate file-backed pages across security domains, including pages belonging to system executables, dynamic libraries, the dynamic loader, and other runtime dependencies. Maintaining private copies of these pages increases physical memory consumption, duplicates page-cache entries, and reduces instruction-cache sharing. 

A second class of countermeasures isolates cache state across security domains. CATalyst~\cite{liu2016catalyst} uses Intel Cache Allocation Technology (CAT) \cite{cat} to divide the LLC into secure and non-secure regions and pins security-sensitive code and data pages in the secure region. DAWG~\cite{kiriansky2018dawg} isolates cache hits, misses, and metadata updates across domains, thereby preventing an attacker from observing a cache line loaded by the victim even when the underlying memory is shared. However, these approaches inevitably incur varying degrees of cache performance degradation.

\section{Discussion}
% \shinan{lead-in is needed; also, limitation about non-token-level recovery}
In this section, we discuss the impact of multi-agent execution and the applicability to other types of agents.
\subsection{Multi-agent Noise}
When multiple agents execute different tasks concurrently, their tool-invocation workflows become interleaved, which degrades the effectiveness of both the \emph{Task Profiling Attack} and the \emph{Trigger Localization Attack}. Fortunately, certain agent behaviors enable the attacker to determine whether multiple agents are running simultaneously.

In our experiments, we observe that Codex consistently exhibits a characteristic tool-invocation sequence \texttt{cat → set → awk → sed → wc → tr → sed → wc → tr}  at the beginning of each trace. By examining the Codex source code \cite{srccodex}, we find that this pattern originates from the shell snapshot procedure performed during session initialization. Specifically, the first cat invocation reads the login profile required by \texttt{bash -lc}, while \texttt{set → awk} and the two subsequent \texttt{sed → wc → tr} invocations process shell options and exported variables to construct the snapshot. Monitoring these additional tools could therefore provide an even more reliable indicator of the shell initialization phase. Claude Code exhibits similar initialization behavior. In addition to taking a shell snapshot, Claude Code also performs a Git snapshot during initialization, which appears in our traces as four consecutive \texttt{git} invocations. The adversary can also monitor rises and falls in CPU utilization to estimate the agent’s execution window and use the sequence of tool invocations during initialization to assess whether those changes are attributable to the agent. These characteristic patterns can help the attacker determine whether multiple agents are running concurrently and, consequently, whether a collected trace may be contaminated by multi-agent interference. We leave the problem of performing the \emph{Task Profiling Attack} and \emph{Trigger Localization Attack} directly on such interleaved multi-agent traces to future work.

\subsection{Other Types of Agents}
For different types of agents and tasks, the set of monitored tools should be adapted accordingly. For the computer-engineering tasks considered in this work, monitoring OS-level tools is sufficient to capture the relevant agent actions. Other agent categories beyond our scope, such as Web agents \cite{deng2023mind2web,yang2025agentoccam}, Productivity agents \cite{boisvert2024workarena++}, and Business agents \cite{yao2024tau,barres2025tau}, may require monitoring the corresponding tools used in their execution environments. For example, attacking Web agents may require monitoring browser-oriented tools such as navigate, click, and screenshot to reconstruct the agent’s web interactions and thereby infer privacy-sensitive information about the user. 

Some agent implementations require a VM-based environment to host the agent, such as Devin \cite{devin}, Cursor Cloud Agents \cite{cursorcloud} and Google Jules \cite{Jules}. Since these agents do not execute locally, they cannot perform tasks that require interaction with the user’s local environment, such as configuring environment of user’s system. As most practical agents are not deployed in such a manner, these VM-hosted agents are out of scope for our attack.
\section{Conclusion}
In this work, we systematically analyzed the cache side-channel risks introduced by local tool execution in LLM agents. Our evaluation across three hardware platforms shows that \emph{Flush}+\emph{Reload} enables accurate, real-time monitoring of agentic tool invocations, while \emph{Prime}+\emph{Probe} provides cross-VM capability at the cost of substantial monitoring overhead and cache-set aliasing. Building on these findings, we presented the \emph{Tool Recovery Attack}, which reconstructs tool-execution workflows across 11 LLMs in Codex and Claude Code, with the lowest F1 score exceeding 93\%. We further showed that the recovered traces enable attack scenarios. The \emph{Task Profiling Attack} recovers the user’s task topic with over 87\% accuracy, while the \emph{Trigger Localization Attack} reduces the overhead of launching fine-grained attacks to less than 20\% of the original level while maintaining recall above 96\%.

\appendix
\renewcommand{\thesubsectiondis}{\thesubsection.}
\subsection{Details of CLLA, DAA and OSA}
\label{appendix:argdetail}
In this section, we describe the detailed implementations of CLLA, DAA, and OSA. CLLA is presented in Algorithm \ref{alg:CLA}, DAA in Algorithm \ref{alg:da}, and OSA in Algorithm \ref{alg:offset_search}.

\subsection{Handle Large Size of SF Sets in \emph{Prime}+\emph{Probe}}
\label{appedix:hlsp+p}
As discussed in Section \ref{subsection:p+p}, monitoring all SF sets of a modern server-class CPU is impractical. To make the \emph{Prime}+\emph{Probe} practical, it is therefore necessary to identify the subset of SF sets that are actively exercised by the victim and focus monitoring on those sets. To achieve this, we adopt a method similar to prior work \cite{zhao2024last}: we collect short memory-access traces for candidate SF sets and estimate their Power Spectral Density (PSD) to determine whether they are influenced by victim activity. PSD transforms time-domain signals into the frequency domain \cite{welch1967use}. Unlike background noise, victim-induced memory accesses often exhibit periodic patterns, which appear as distinct frequency peaks in the PSD of the corresponding traces.

In our experiments, we limit the adversary to 16 CPU cores. Accordingly, we select 64 victim-related SF sets for monitoring, resulting in a workload of four SF sets per core. The attack follows a cyclic process of \emph{Prime} $\rightarrow$ \emph{Wait} $\rightarrow$ \emph{Probe}. During the \emph{Wait} phase, the CPU core is effectively available for executing the \emph{Prime}/\emph{Probe} phases of the other three assigned SF sets. In other words, the waiting time is equivalent to the time of three \emph{Prime}/\emph{Probe} operations.

%We will show in later experiments that this configuration represents a reasonable trade-off. Increasing the number of monitored SF sets would provide broader coverage. However, it also prolongs the waiting interval. An excessively long waiting window may include multiple victim accesses, which cannot be distinguished within a single \emph{Prime}+\emph{Probe} iteration, thereby leading to information loss.

\subsection{Task Categories in \emph{Task Profiling Attack}}
\label{appendix:taxonomy}
Detailed descriptions of the nine task categories used in \emph{Task Profiling Attack} evaluation are provided in Table \ref{table:taxonomy}.

\begin{algorithm}[t]
\caption{Cache Line Localization Algorithm}
\small
\label{alg:CLA}
\begin{algorithmic}[1]
\Require Binary program $B$
\Ensure Selected mandatory cacheline $c^*$

\State Load $B$ and build a normalized control-flow graph $CFG$
\State $e \gets$ entry node of $CFG$
\State $G \gets$ reachable CFG subgraph starting from $e$

\State Condense strongly connected components of $G$ into an acyclic graph $G_{\mathrm{SCC}}$
%\State Condense strongly connected components of $G$ into a DAG
\State Compute entry-to-terminal path counts on the $G_{\mathrm{SCC}}$

\State $\mathcal{C} \gets \emptyset$
\ForAll{node $v \in G$}
    \State Disassemble the basic block of $v$
    \ForAll{instruction address $a$ in $v$}
        \State $c \gets a - (a \bmod Cacheline\_size)$
        \State Add $v$ to the node set of cacheline $c$
        \State Increment the instruction count of $c$ in $v$
        \State Add $c$ to $\mathcal{C}$
    \EndFor
\EndFor

\State $\mathcal{S} \gets \emptyset$

\ForAll{cacheline $c \in \mathcal{C}$}
    \If{there exists an entry-to-terminal path avoiding $c$}
        \State \textbf{continue}
    \EndIf

    \State $loop\_insts \gets$ number of instructions of $c$ in loops
    \State $total\_visits \gets$ total visits of $c$ over all paths
    \State $avg\_visits \gets total\_visits / total\_paths$

    \State Add $(c, loop\_insts, avg\_visits)$ to $\mathcal{S}$
\EndFor

\If{$\mathcal{S} = \emptyset$}
    \State \Return failure
\EndIf

\State $c^* \gets \arg\min\limits_{c \in \mathcal{S}}
    (loop\_insts, avg\_visits, c)$

\State \Return $c^*$
\end{algorithmic}
\end{algorithm}

\begin{table*}[t]
\centering
\footnotesize
\caption{Representative tasks used in the Trigger Localization Attack datasets.}
\label{tab:trigger_task_examples}
\setlength{\tabcolsep}{3pt}
\begin{tabular}{p{0.13\textwidth} p{0.37\textwidth} p{0.29\textwidth} p{0.16\textwidth}}
\toprule
\textbf{Task type} &
\textbf{Task scenario} &
\textbf{Input and output assets} &
\textbf{Representative aspect} \\
\midrule
Encryption &
\texttt{security\_crypto\_100\_e48}: encrypt an audit log with GPG for a
specified recipient before the log is archived or transferred off-host. &
Inputs:\texttt{ /app/audit.log}, \texttt{/app/recipient\_public.asc}. 

Output:
\texttt{/app/audit.log.gpg}. &
Recipient-specific audit-log protection \\
\addlinespace
Decryption &
\texttt{security\_crypto\_100\_d12}: decrypt an AES-256-CBC-protected SQL
backup as preparation for database recovery or migration. &
Inputs:\texttt{ /app/db\_backup.sql.enc}, \texttt{/app/aes.key},
\texttt{/app/aes.iv}. 

Output: \texttt{/app/db\_backup.dec.sql}. &
Database recovery from encrypted backups \\
\addlinespace
Inference &
\texttt{model\_trigger\_100\_i09}: query a local vLLM server through an
OpenAI-compatible streaming chat-completions endpoint and reconstruct the
assistant response. This models deployed local LLM serving. &
Input:\texttt{ /app/chat\_request.json}. 

Output:
\texttt{/app/vllm\_stream.txt}. &
OpenAI-compatible local LLM serving \\
\addlinespace
Training &
\texttt{model\_trigger\_100\_t21}: fine-tune a local LLM with a PEFT
DoRA adapter using SFT for a small number of optimizer steps. This models
parameter-efficient domain adaptation before deployment. &
Inputs:\texttt{ /app/models/llm-base}, \texttt{/app/domain\_sft.jsonl}.

Output: \texttt{/app/outputs/dora\_domain}. &
Parameter-efficient domain adaptation \\
\bottomrule
\end{tabular}
\vspace{-0.5cm}
\end{table*}

\subsection{Details of Attack Methodologies}

\subsubsection{Task Profiling Attack}
\label{appendix:taskprofiling}

\textbf{Semantic Tool Representation.}
%\label{sububsection:ToolRepresentation}
A straightforward representation assigns an independently learned vector to each tool. However, such a representation ignores the semantic relationships among tools. For example, tools for reading and searching files may play similar roles even though they have different names. To incorporate this information, we initialize the tool representations using a pretrained sentence-embedding model \cite{awasthy2026graniteembeddingmultilingualr2}.

Let $\mathcal{V}$ denote the tool vocabulary constructed from the training dataset. For every tool $t \in \mathcal{V}$, it is encoded as

\begin{equation}
    \mathbf{e}_{t} = \Phi\left(s_t\right) \in \mathbb{R}^{d_e},
\end{equation}

where $s_t$ is the textual name of tool $t$, $\Phi(\cdot)$ is the pretrained semantic encoder, and $d_e$ is the embedding dimension. We further normalize the embeddings as follows
\begin{equation}
    \mathbf{e}_{t} \leftarrow
    \frac{\mathbf{e}_{t}}{\left\lVert\mathbf{e}_{t}\right\rVert_2}.
\end{equation}

\textbf{Bidirectional Trace Encoder.}
%\label{subsubsection:bidirection}
Tool traces exhibit strong sequential dependencies. For instance, two task categories may invoke an overlapping set of tools but arrange them in different orders. We therefore employ a bidirectional gated recurrent unit (BiGRU) to capture both preceding and subsequent tool-invocation context.

Each invocation is represented by its semantic tool embedding and two timing features. For invocation $j$ in trace $i$, let $u_{i,j}=a_{i,j}-a_{i,1}$ denote elapsed time since the first recorded call. Let $\delta_{i,j}$ denote the gap since the preceding call. Both features are expressed in milliseconds. We apply a nonnegative clamp and a logarithmic transform:
\begin{equation}
    \mathbf{r}_{i,j} =
    \begin{bmatrix}
        \log\!\left(1+\max(u_{i,j},0)\right) \\
        \log\!\left(1+\max(\delta_{i,j},0)\right)
    \end{bmatrix}.
\end{equation}
The transformed features are normalized componentwise:
\begin{equation}
    \bar{\mathbf{r}}_{i,j}
    =
    (\mathbf{r}_{i,j}-\boldsymbol{\mu})
    \oslash\boldsymbol{\sigma}.
\end{equation}
Here, $\oslash$ denotes elementwise division. The normalization means $\boldsymbol{\mu}$ and standard deviations $\boldsymbol{\sigma}$ are fitted on training traces and remain fixed for all held-out traces.

A two-layer MLP encodes the normalized features:
\begin{equation}
    \mathbf{g}_{i,j}
    =
    \mathbf{W}_{T,2}\operatorname{ReLU}
    \left(
        \mathbf{W}_{T,1}\bar{\mathbf{r}}_{i,j}
        +\mathbf{b}_{T,1}
    \right)
    +\mathbf{b}_{T,2}.
\end{equation}
The default time encoder has layer widths $2\rightarrow16\rightarrow16$, with ReLU between the two affine layers. We concatenate its output with the semantic tool embedding:
\begin{equation}
    \mathbf{v}_{i,j}
    =
    \mathbf{e}_{t_{i,j}}\mathbin{\|}\mathbf{g}_{i,j}
    \in\mathbb{R}^{d_e+d_T},
\end{equation}
where $d_T=16$ by default and $\|$ denotes vector concatenation. The time encoder is trained jointly with the trace classifier. For a trace $\mathcal{T}_i$, the resulting input sequence is
\begin{equation}
    \mathbf{X}_i =
    \left(
        \mathbf{v}_{i,1},
        \mathbf{v}_{i,2},
        \ldots,
        \mathbf{v}_{i,L_i}
    \right).
\end{equation}
The forward and backward GRUs process the sequence in opposite directions:
\begin{align}
    \overrightarrow{\mathbf{h}}_{i,j}
    &=
    \operatorname{GRU}_{f}
    \left(
        \mathbf{v}_{i,j},
        \overrightarrow{\mathbf{h}}_{i,j-1}
    \right), \\
    \overleftarrow{\mathbf{h}}_{i,j}
    &=
    \operatorname{GRU}_{b}
    \left(
        \mathbf{v}_{i,j},
        \overleftarrow{\mathbf{h}}_{i,j+1}
    \right).
\end{align}
We concatenate the final states from the two directions to construct a fixed-length representation of the complete trace:
\begin{equation}
    \mathbf{z}_i =
    \overrightarrow{\mathbf{h}}_{i,L_i}
    \mathbin{\|}
    \overleftarrow{\mathbf{h}}_{i,1}
    \in \mathbb{R}^{2d_h},
\end{equation}
$d_h$ denotes the hidden dimension of each GRU direction.

Because task profiling is conducted offline after collecting the complete trace, the attacker can use both preceding and subsequent tool invocations and their timing.

\textbf{Task Classification.}
The trace representation is passed to a two-layer multilayer perceptron:

\begin{align}
    \mathbf{q}_i
    &=
    \operatorname{ReLU}
    \left(
        \mathbf{W}_1 \mathbf{z}_i + \mathbf{b}_1
    \right), \\
    \mathbf{o}_i
    &=
    \mathbf{W}_2
    \operatorname{Dropout}\left(\mathbf{q}_i\right)
    + \mathbf{b}_2,
\end{align}
where $\mathbf{o}_i \in \mathbb{R}^{C}$ contains the classification logits. The probability assigned to category $c$ is

\begin{equation}
    P\left(y_i=c \mid \mathcal{T}_i\right)
    =
    \frac{\exp(o_{i,c})}
    {\sum_{k=1}^{C}\exp(o_{i,k})}.
\end{equation}

Given a training set containing $N$ traces, we optimize the categorical
cross-entropy loss:
\begin{equation}
    \mathcal{L}_{\mathrm{CE}}
    =
    -\frac{1}{N}
    \sum_{i=1}^{N}
    \log P\left(y_i \mid \mathcal{T}_i\right).
\end{equation}
The resulting classifier jointly encodes tool semantics, invocation order, and call timing to infer the task category.

\begin{algorithm}[t]
\caption{Deduplication Attack Algorithm}
\small
\label{alg:da}
\begin{algorithmic}[1]
\Require Address $addr$, threshold $\tau$, unlock parameter $U$
\Ensure Detected event count $call\_count$

\State $call\_count \gets 0$
\State $locked \gets \textbf{false}$
\State $miss\_streak \gets 0$

\While{the monitoring phase is active}
    \State \Call{Flush}{$addr$}
    \State \Call{Wait}{}
    \State $latency \gets$ \Call{Reload}{$addr$}

    \If{$latency < \tau$}
        \If{\textbf{not} $locked$}
            \State $call\_count \gets call\_count + 1$
            \State $locked \gets \textbf{true}$
        \EndIf
    \EndIf

    \If{$locked$}
        \If{$latency \geq \tau$}
            \State $miss\_streak \gets miss\_streak + 1$
            \If{$miss\_streak \geq U$}
                \State $locked \gets \textbf{false}$
                \State $miss\_streak \gets 0$
            \EndIf
        \Else
            \State $miss\_streak \gets 0$
        \EndIf
    \EndIf
\EndWhile

\State \Return $call\_count$
\end{algorithmic}
\end{algorithm}

\subsubsection{Trigger Localization Attack}
\label{appendix:triggerLocalization}
\textbf{Online Sequence Labeler.}
We encode each tool using the semantic embedding scheme described in Section~\ref{appendix:taskprofiling}. The timing features and time-encoding MLP follow the construction in Section~\ref{appendix:taskprofiling}. For this attack, normalization statistics are fitted only on the localization training traces and remain fixed during model selection, calibration, and testing. The time encoder is trained jointly with the localization model.

For the current invocation, elapsed time is measured from the first recorded call. Both features use only invocation timestamps observed through position $t$. We concatenate their $\mathbf{g}_t$ with the semantic tool embedding to form
\begin{equation}
    \mathbf{v}_t
    =
    \mathbf{e}_{x_t}\mathbin{\|}\mathbf{g}_t.
\end{equation}
Unlike the task profiling attack in Section~\ref{appendix:taskprofiling}, trigger localization uses a unidirectional GRU. Task profiling operates on a complete trace, whereas localization predicts from the observed prefix. The causal GRU updates its hidden state as
\begin{equation}
    \mathbf{h}_t
    =
    \operatorname{GRU}
    \left(\mathbf{v}_t,\mathbf{h}_{t-1}\right),
\end{equation}
where $\mathbf{h}_{t-1}$ summarizes the previously observed tools and timestamps $(x_{1:t-1},a_{1:t-1})$. Consequently, $\mathbf{h}_t$ depends only on $(x_{1:t},a_{1:t})$. This update can be applied one invocation at a time while retaining the hidden state.

A dropout layer and a linear classification head are applied independently at every trace position:
\begin{equation}
    \mathbf{o}_t
    =
    \mathbf{W}\operatorname{Dropout}(\mathbf{h}_t)
    +\mathbf{b},
\end{equation}
where $\mathbf{o}_t \in \mathbb{R}^{2}$ contains the logits for the two state labels. We convert the logits into probabilities using
\begin{equation}
    p_{t,c}
    =
    P(y_t=c\mid x_{1:t},a_{1:t})
    =
    \frac{\exp(o_{t,c})}
    {\sum_{k\in\{0,1\}}\exp(o_{t,k})}.
\end{equation}
An invocation is predicted as sensitive when $p_{t,1}\geq\tau$.

\textbf{Class-Imbalance-Aware Training.}
False negatives may cause the attacker to miss sensitive invocations. We therefore train with class-weighted cross-entropy and select the positive-class weight under the recall-constrained objective.

Let $n_c$ denote the number of valid training positions belonging to class $c$. The class weights are
\begin{equation}
    w_0=1,
    \qquad
    w_1=\alpha,
    \qquad
    \alpha>0.
\end{equation}
By default, we search $\alpha\in\{1,2,4,8,16\}\cup\{n_0/n_1\}$. The imbalance ratio $n_0/n_1$ is computed only from training labels. Each candidate weight is used to train a separate model.

For a $\mathcal{B}$, the weighted cross-entropy objective is
\begin{equation}
    \mathcal{L}_{\mathrm{WCE}}(\mathcal{B})
    =
    -\frac{1}{Z_{\mathcal{B}}}
    \sum_{i\in\mathcal{B}}
    \sum_{t=1}^{L_i}
    w_{y_{i,t}}\log p_{i,t,y_{i,t}},
\end{equation}
where $L_i$ is the number of valid positions in trace $i$, $p_{i,t,y_{i,t}}$ is the predicted probability of its ground-truth label, and
\begin{equation}
    Z_{\mathcal{B}}
    =
    \sum_{i\in\mathcal{B}}L_i.
\end{equation}
%The loss is averaged over valid positions; its denominator is the number of valid positions rather than the sum of their class weights. Padding positions are excluded from both the loss and the denominator. By increasing the contribution of sensitive positions during training, the weighted objective discourages the model from achieving deceptively high accuracy by predicting most invocations as non-sensitive.

\begin{table}[tb]
    \begin{center}
    \caption{Taxonomy of \emph{Task Profiling Attack}.}
    \small
    \label{table:taxonomy}
    \scalebox{0.9}{
        \begin{threeparttable}
        \begin{tabular}{c|c|c}
        \toprule[2pt]
           & Categories & Description\\
         \hline
         $T_1$ & Data Science & Data processing and statistic\\
         \hline
         $T_2$ & Debugging & Fixing Code Errors\\
         \hline
         $T_3$ & File Operations & Finding/modifying files\\
         \hline
         $T_4$ & Games & Solving chess problems\\
         \hline
         $T_5$ & Model Training & Model-related tasks\\
         \hline
         $T_6$ & Scientific Computing & Science-related tasks\\
         \hline
         $T_7$ & Security & Security-related tasks\\
         \hline
         $T_8$ & Software Engineering & Coding, build-tools\\
         \hline
         $T_9$ & System Administration & Administration Operations\\
        \bottomrule[2pt]
        \end{tabular}
        \end{threeparttable}}
    \end{center}
    \vspace{-0.5cm}
\end{table}

\begin{algorithm}[t]
\caption{Offset Search Algorithm}
\small
\label{alg:offset_search}
\begin{algorithmic}[1]
\Require Base address $base$, cache hit test rounds $R$
\Ensure Detected target offset $offset$

\State $best\_score \gets -\infty$
\State $best\_offset \gets 0$

\For{$offset \gets 0$ to $Lib\_Size$ step $Cache\_Line$}
    \State $target\_hits \gets 0$
    \State $other\_hits \gets 0$

    \For{$i \gets 1$ to $R$}
        \State \Call{Flush}{$base + offset$}
        \State \Call{Execute}{target instruction argument}
        \State $latency \gets$ \Call{Reload}{$base + offset$}
        \If{$latency < \tau$}
            \State $target\_hits \gets target\_hits + 1$
        \EndIf

        \ForAll{other instruction arguments}
            \State \Call{Flush}{$base + offset$}
            \State \Call{Execute}{current argument}
            \State $latency \gets$ \Call{Reload}{$base + offset$}
            \If{$latency < \tau$}
                \State $other\_hits \gets other\_hits + 1$
            \EndIf
        \EndFor
    \EndFor

    \State $score \gets target\_hits - other\_hits$

    \If{$score = R$}
        \State \Return $offset$
    \EndIf

    \If{$score > best\_score$}
        \State $best\_score \gets score$
        \State $best\_offset \gets offset$
    \EndIf
\EndFor

\State \Return $best\_offset$
\end{algorithmic}
\end{algorithm}

\subsection{Representative Trigger-Localization Tasks}
\label{appendix:trigger_examples}

\textbf{Dataset scale.}
For task profiling, we use 80 tasks from Terminal-Bench~1 and 89 tasks from
Terminal-Bench~2. The evaluation covers five GPT variants, four Claude
variants, and two GLM variants. We collect up to multiple runs for each task and model combination. %After trace quality
%filtering, the task profiling dataset contains 4000 traces. 
For trigger
localization, we construct two balanced suites with 100 tasks each. The key extraction suite contains
50 encryption tasks and 50 decryption tasks, while the model extraction suite
contains 50 inference and 50 training tasks.% Each task is run three times on the same 11 models, which produces 3000 traces per suite and 6000
%trigger localization traces in total.

\textbf{Active labeling.}
We label sensitive steps with an active review procedure focused on likely
errors, instead of manually inspecting every trace. First, an automatic annotator generates
candidate labels from task metadata and execution evidence. In the security
suite, the invocation must also target the requested output, while setup, key inspection, and validation operations receive negative labels.
In the model suite, the annotator combines task specific expected triggers with instrumented framework operation timestamps and endpoint records. This evidence identifies the
invocation that performs inference or training, and separates it from the
surrounding steps. After each pass, we prioritize traces with no positive step
or unusually many positive steps. We also prioritize traces with fallback output
attribution, conflicts between semantic and runtime evidence, or unmatched
recorded operations. We inspect these cases, refine the matching rules, and
relabel the corpus. When the agent never executes the target operation, a trace
retains all zero labels. This active review loop focuses annotation effort on
ambiguous traces, while preserving the binary step labels used to train the
localization model.

Table~\ref{tab:trigger_task_examples} presents four representative tasks, taken
from our two Trigger Localization Attack datasets. The four workflows are GPG
protection of audit logs, recovery of an AES encrypted database backup,
OpenAI compatible local LLM serving, and parameter efficient domain adaptation.

The examples cover public key encryption and symmetric decryption, as well as
streaming model inference and fine tuning before deployment. Their operational
settings cover audit retention and database recovery, as well as deployed model
serving and domain adaptation.

\end{document}